\documentclass[oldversion]{aa} 
\usepackage{graphicx}
\usepackage{aalongtable}
\usepackage{txfonts}
\usepackage{rotating}
\usepackage{lscape}
\newcommand{\gsim}{\;\lower.6ex\hbox{$\sim$}\kern-7.75pt\raise.65ex\hbox{$>$}\;}
\newcommand{\lsim}{\;\lower.6ex\hbox{$\sim$}\kern-7.75pt\raise.65ex\hbox{$<$}\;}

\begin{document}

\title{Empirical estimates of the Na-O anti-correlation in 95 Galactic 
globular clusters
} 

\author{
Eugenio Carretta\inst{1}
}

\authorrunning{Eugenio Carretta}
\titlerunning{Empirical estimates of Na-O anti-correlation}

\offprints{E. Carretta, eugenio.carretta@inaf.it}

\institute{
INAF-OAS Osservatorio di Astrofisica e Scienza
dello Spazio di Bologna via Gobetti 93/3, I-40129
 Bologna, Italy}

\date{}

\abstract{Large star-to-star abundance variations are direct evidence of 
multiple stellar populations in Galactic globular clusters (GCs). The main and
most widespread chemical signature is the anti-correlation of the stellar Na and O
abundances. The interquartile range (IQR) of the [O/Na] ratio is well suited to
quantifying the extent of the anti-correlation and to probe its links to global
cluster parameters. However, since it is quite time consuming to obtain precise
abundances  from spectroscopy for large samples of stars in GCs, here we show
empirical calibrations of IQR[O/Na] based on the O, Na abundances homogeneously 
derived from more than 2000 red giants in 22 GCs in our FLAMES survey. We find
a statistically robust bivariate correlation of IQR as a function of the total
luminosity (a proxy for mass) and cluster concentration $c$. Calibrated and observed
values lie along the identity line when a term accounting for the horizontal
branch (HB) morphology is added to the calibration, from which we obtained
empirical values for 95 GCs. Spreads in proton-capture elements O and Na are found
for all GCs in the luminosity range from $M_V=-3.76$ to $M_V=-9.98$. This
calibration reproduces in a self-consistent picture the link of abundance
variations in light elements with the He enhancements and its effect on the
stellar distribution on the HB. We show that the spreads in light elements seem already
to be dependent on the initial GC masses. The dependence of IQR on 
structural parameters stems from the well known correlation between $c$ and
$M_V$, which is likely to be of primordial origin. Empirical estimates can be used to
extend our investigation of multiple stellar populations to GCs in external
galaxies, up to M31, where even integrated light spectroscopy may currently
provide only a hint of such a phenomenon.
}

\maketitle

\section{Introduction}
The dawn of observations of multiple stellar populations in Galactic globular
clusters (GCs) dates back to more than 30 years ago. Since then, a major effort
has been devoted to quantifying the amount  of the stellar component whose composition
deviates from the typical pollution  given by core-collapse supernovae (SNe) in
a metal-poor environment. 

The first attempts concerned the dichotomy of CN and CH bandstrengths observed
in GC stars (primarily giants, see the focus review by Smith 1987). The ratios
of CN-weak to CN-strong stars were used as fingerprints of the star to star
abundance variations in GCs and to investigate possible links to global cluster
parameters (see e.g. Norris 1987, and, more recently Kayser et al. 2008, Pancino
et al. 2010). This approach essentially discriminates two main components: first
generation (FG) stars,  with the typical chemical pattern of metal-poor halo
stars, and second generation (SG) stars, whose composition was very likely
altered by the nuclearly processed matter ejected from massive stars of the
first component. This, in short, is the classical scenario summarized, together
with its history, drawbacks and open issues, in a number of reviews with all the
necessary references (Kraft 1994, Gratton et al. 2004, 2012, Bastian and Lardo
2018).

An obvious advantage of CH/CN-based indicators is that the method can be easily
extended to large samples by using suitable photometric bandpasses sensitive to
C and N absorption features (see e.g. Briley 1997, Briley et al. 2001, Milone et
al. 2012a, Monelli et al. 2013). Among the cons, we mention  the warning put
forward by G.H. Smith and collaborators (see Smith et al. 2013, Smith 2015),
that probably photometry and spectroscopy may not see exactly the same things,
most photometric indexes being `blind' to the alterations in the abundances of
heavier elements.  As a consequence, it is more difficult to quantify the
contributions when higher ranges of temperatures are involved in the
proton-capture  reactions of the complete CNO cycle (Denisenkov and Denisenkova
1989, Langer et al. 1993) responsible for alterations in species such as O and Na, not
to mention Mg and Al.

A typical example is provided by the anti-correlation of Na with O abundances in
GC stars discovered by the Lick-Texas group about 30 years ago (see Kraft 1994, Gratton et al.
2004 for references). 
As large samples of stars in several GCs became routinely available thanks to
efficient, high multiplex spectrographs, Carretta (2006) proposed to use the
interquartile range IQR of the [O/Na] distribution to quantify the extent of
this main chemical signature of multiple populations in GCs. Being defined as
the middle 50\% of a data sample, the IQR is a robust indicator, insensitive to
outliers, not very sensitive to small number statistics and upper limits.
The IQR is driven by the fraction of stars where the changes in composition are
more evident (Carretta et al. 2010a). The IQR[O/Na] seems to be a good tool to
probe the interplay between the internally processed matter and the leftover
pristine gas resulting in the formation of SG stars. 

\begin{figure}
\centering
\includegraphics[bb=33 176 295 701, clip, scale=0.52]{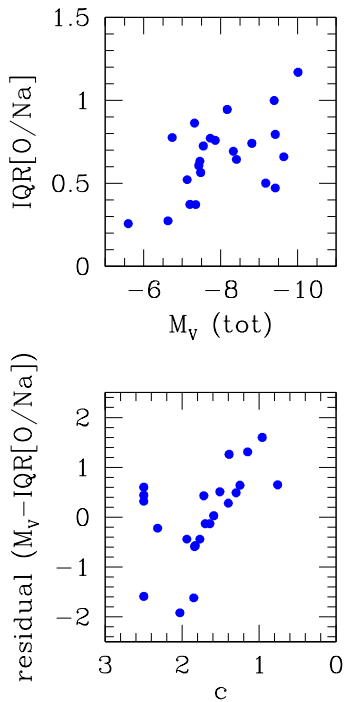}
\caption{IQR[O/Na] for Galactic GCs as a function of the total absolute
magnitude $M_V$ (upper panel), and residual around this relation as a function
of the cluster concentration $c$ (lower panel). Both panels are reproduced from
Carretta et al. (2014a).}
\label{f:iqrres48}
\end{figure}

However, despite the modern instrumentation, this approach may be time consuming,
if very fertile. During our FLAMES survey devoted to the link between the
Na-O anti-correlation and horizontal branch (HB) morphology  (Carretta et al.
2006) we invested about 12 years to collect homogeneous  abundances for more
than 2500 red giant branch (RGB) stars in 25 GCs. From this unprecedented
database, we found that IQR[O/Na] is well correlated with the present-day total
cluster mass, as expressed for example, by the proxy of total absolute magnitude $M_V$. 
Moreover, the larger the extent of the anti-correlation, the higher the 
maximum temperature reached by stars populating the HB in the GCs (Carretta
et al. 2007a) is. The IQR[O/Na]-$M_V$ correlation, although very significant, shows
some  scatter, and in  Carretta et al. (2014a) we uncovered that the residuals
around this relation are not randomly distributed, but they are correlated very 
well with the cluster concentration parameter $c$, once the core-collapse GCs
are excluded (see Fig.~\ref{f:iqrres48}).

These relations are so statistically robust that we can use almost all
the GCs in our FLAMES survey as calibrators and provide a multivariate relation
to empirically estimate the extent of the Na-O anti-correlation in many more
objects, even without spectroscopic observations of individual stars. These
empirical estimates are then possible even for small or distant GCs, poorly
studied  not only with high-resolution spectroscopy, but also neglected by most
photometric surveys. 

Moreover, beside their use as probes of multiple populations in never-before-explored 
GCs, the derived relations may hide in plain sight deeper physical
meanings. The present-day total mass of a GC is nothing but a reflection of its 
initial mass, as combined with internal dynamical processes and external 
dynamical evolution due to interactions with the host Galaxy in almost a Hubble 
time. In a sense, these processes can be summarized in the parameter $c$ (e.g.
Murray and Lin 1992, Djorgovski 1991). The elusive cluster-to-cluster spreads in
He are accounted for by a parameter describing the HB morphology. These are 
the ingredients of the new calibration proposed in the present work.

This new tool can be applied to a large number of GCs belonging to the different
Galactic sub-populations, providing empirical estimates of the Na-O
anti-correlation for about 80\% of the whole GC system of the Milky Way.
Moreover, the range of application extends at least up to galactic systems
where HB stars can be currently resolved in GCs, such as the M~31 galaxy. A better
and statistically robust understanding of the links between the global cluster
properties and the multiple population  phanomenon is then possible.

The paper is organized as follows: we describe in Section 2 how our empirical 
calibration is derived and how it can be improved. Sanity checks, the
application to the Galactic GCs, and the link with the properties of different 
globular cluster sub-populations are illustrated in Section 3. These results are
discussed in Section 4, with emphasis on the link between initial and
present-day GC masses. A summary is given in Section 5.

\section{Calibration of the Na-O extension in GCs}

The main outcome of our FLAMES survey is represented by homogeneous Fe abundances for
more than 2500 RGB stars and Na, O abundances for most of them, allowing us to
derive the observed values of IQR[O/Na]. The most updated versions of the 
IQR[O/Na]-$M_V$ correlation and the one of its residual as a function of
concentration $c$ is given in Bragaglia et al. (2017, their Fig. 5).

\subsection{First calibration}

The first calibration we may think of is then simply an update of the bivariate
analysis shown in Carretta et al. (2014a), where the derived IQR can be 
calibrated as a combination of the cluster total absolute magnitude $M_V$ and
concentration $c$. The updated sample simply consists in adding the value for
NGC~6139 (Bragaglia et al. 2015) and the recent reanalysis of NGC~6388 with a
larger sample of stars (Carretta and Bragaglia 2018). 

Among the calibrators, three core-collapse GCs present in  our
FLAMES survey (NGC~6752, NGC~6397, and NGC~7099) as well as NGC~6535  (Bragaglia
et al. 2017) are not considered. The last one is not listed as core-collapse in the Harris (1996, 2010
on line edition; hereinafter H10) catalogue. However, its position in the
residual-$c$ plane appears to be clearly discrepant with respect to the other
GCs, indicating some peculiarity for this cluster, as discussed in Bragaglia et
al. (2017). NGC~6535 is a GC with an inverted mass function (Halford and
Zaritsky 2015), with a current $M_V=-4.75$ (H10) and a large IQR[O/Na].
This GC was severely depleted of its low mass stars by interaction with the
Galaxy. For NGC~6535 to lie on the relation defined by the other GCs we would
need to assume a luminosity at least 2 mag brighter than the current one. This
uncertainty makes this object not well suited as a good calibrator.

\begin{table*}
\centering
\caption{Relevant parameters for calibrating GCs from our FLAMES survey}
\begin{tabular}{lcccccrl}
\hline
NAME            &  $M_V$  &  c &  IQR[O/Na] & [Fe/H]&   t &   HBR & References for \\
                &  H10  & H10&   obs.     & H10  &        &       & IQR[O/Na] \\
\hline

NGC0104  47Tuc  &  -9.42 & 2.07 &   0.472  &   -0.72  &  0 &  -0.99 & Carretta et al. (2009a,b) \\
NGC0288         &  -6.75 & 0.99 &   0.776  &   -1.32  &  1 &   0.98 & Carretta et al. (2009a,b) \\
NGC0362         &  -8.43 & 1.76 &   0.644  &   -1.26  &  1 &  -0.87 & Carretta et al. (2013) \\
NGC1851         &  -8.33 & 1.86 &   0.693  &   -1.18  &  2 &  -0.32 & Carretta et al. (2011) \\
NGC1904  M79    &  -7.86 & 1.70 &   0.759  &   -1.60  &  2 &   0.89 & Carretta et al. (2009a,b) \\
NGC2808         &  -9.39 & 1.56 &   0.999  &   -1.14  &  1 &  -0.49 & Carretta (2015) \\
NGC3201         &  -7.45 & 1.29 &   0.634  &   -1.59  &  1 &   0.08 & Carretta et al. (2009a,b) \\
NGC4590  M68    &  -7.37 & 1.41 &   0.372  &   -2.23  &  1 &   0.17 & Carretta et al. (2009a,b) \\
NGC4833         &  -8.17 & 1.25 &   0.945  &   -1.85  &  1 &   0.93 & Carretta et al. (2014a) \\
NGC5904  M5     &  -8.81 & 1.73 &   0.741  &   -1.29  &  1 &   0.31 & Carretta et al. (2009a,b) \\
NGC6093  M80    &  -8.23 & 1.68 &   0.784  &   -1.75  &  1 &   0.93 & Carretta et al. (2015) \\
NGC6121  M4     &  -7.19 & 1.65 &   0.373  &   -1.16  &  1 &  -0.06 & Carretta et al. (2009a,b) \\
NGC6139         &  -8.36 & 1.86 &   0.647  &   -1.65  &  1 &   0.91 & Bragaglia et al. (2015) \\
NGC6171  M107   &  -7.12 & 1.53 &   0.522  &   -1.02  &  0 &  -0.73 & Carretta et al. (2009a,b) \\
NGC6218  M12    &  -7.31 & 1.34 &   0.863  &   -1.37  &  0 &   0.97 & Carretta et al. (2007b) \\
NGC6254  M10    &  -7.48 & 1.38 &   0.565  &   -1.56  &  0 &   0.98 & Carretta et al. (2009a,b) \\
NGC6388         &  -9.41 & 1.75 &   0.644  &   -0.55  &  0 &  -1.00 & Carretta \& Bragaglia (2018) \\
NGC6441         &  -9.63 & 1.74 &   0.660  &   -0.46  &  0 &  -1.00 & Gratton et al. (2006,2007) \\
NGC6715  M54    &  -9.98 & 2.04 &   1.169  &   -1.49  &  3 &   0.54 & Carretta et al. (2010b) \\
NGC6809  M55    &  -7.57 & 0.93 &   0.725  &   -1.94  &  0 &   0.87 & Carretta et al. (2009a,b) \\
NGC6838  M71    &  -5.61 & 1.15 &   0.257  &   -0.78  &  0 &  -1.00 & Carretta et al. (2009a,b) \\
NGC7078  M15    &  -9.19 & 2.29 &   0.501  &   -2.37  &  1 &   0.67 & Carretta et al. (2009a,b) \\
\hline
\end{tabular}
\begin{list}{}{}
\item[(1)] H10: from Harris (1996, 2010 on line edition)
\item[(2)] t: Galactic population (from Carretta et al. 2010a): 0=bulge/disc, 1=inner halo, 2=outer halo,
3=dSph (Sgr)
\item[(3)] HBR: from Mackey and van den Bergh (2005)
\end{list}
\label{t:tabcal}
\end{table*}

The relevant data for the remaining 22 GCs used as calibrators are listed in 
Table~\ref{t:tabcal}. The values for $M_V$, $c$, and total metallicity [Fe/H]
are from H10, the HB index or ratio, HBR (Zinn
1986, Lee 1989, 1990), is from Mackey and van den Bergh (2005), and the observed
IQR[O/Na] values are from the studies whose references are given in the last 
column. For the clusters NGC~1851 (M~54), showing a small (noticeable)
iron spread and the Na-O anti-correlation in each metallicity component, we
adopted the total values as in Fig.~\ref{f:iqrres48}.
For the classification of GCs in Galactic populations (halo, bulge,
disc) we followed Carretta et al. (2010a).

Our first calibration is then:

\begin{eqnarray}
IQR1 = -0.193(\pm 0.049) M_V -0.429(\pm 0.153) c \nonumber \\
-0.221(\pm 0.035)
\end{eqnarray}
with $rms=0.166$ and a Pearson regression coefficient $r_p=0.67$ (22 GCs).
The two-tail probability to be a random result is $p=6.5\times 10^{-4}$.

\subsection{A missing ingredient}

\begin{figure}
\centering
\includegraphics[scale=0.40]{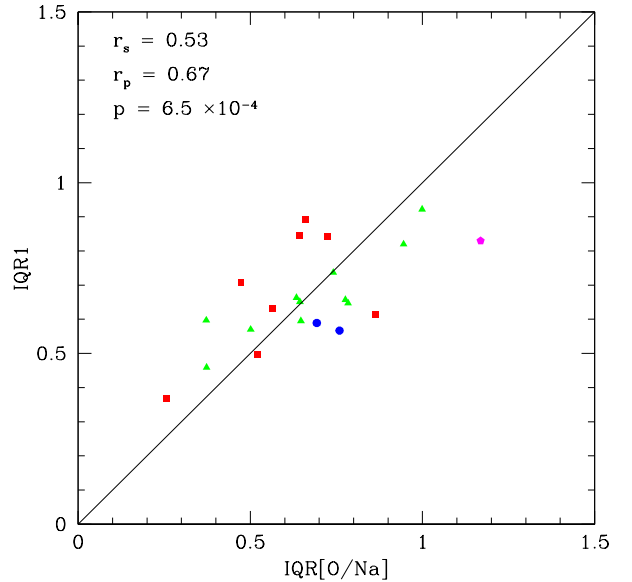}
\caption{Comparison of the IQR1 values from the first calibration (equation 1)
with the observed values IQR[O/Na] for 22 calibrationg GCs. The Spearman rank
correlation ($r_s$) {coefficient, the Pearson linear correlation ($r_p$) 
coefficients and its p-value} are listed in the panel. Different symbols
individuate different Galactic populations (see text).}
\label{f:cfriqr1}
\end{figure}

In Fig.~\ref{f:cfriqr1} the calibrated IQR1 is plotted as a function of the
actually observed values of IQR[O/Na] for our 22 calibrating GCs. Different
symbols, used throughout this paper, individuate GCs of the bulge and/or disc 
(hereafter bulge/disk; filled
squares), inner halo (triangles), and outer halo (circles) components, as well
as those associated to the dwarf spheroidal galaxy Sagittarius, currently disrupting in
our Galaxy (pentagons).

As shown by the Spearman rank correlation coefficient $r_s$ and the Pearson
linear correlation coefficient $r_p$, the relation between IQR1 and IQR[O/Na] is
good, statistically significant at more than 99.9\% level of confidence.
However, the calibrated values deviate from the equality line: something is 
still missing.

\begin{figure*}
\centering
\includegraphics[bb=18 158 590 428, clip, scale=0.82]{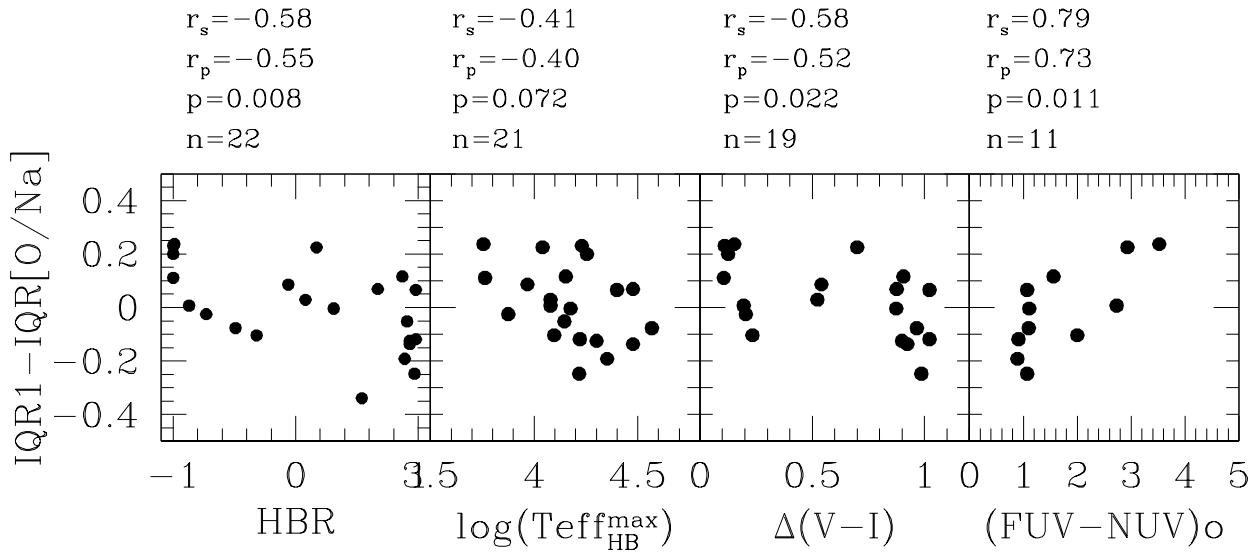}
\caption{Residuals from the calibration of Equation (1) as a function of various
parameters for the HB morphology. Above each panel the Spearman and Pearson
correlation coefficients are listed, as well as the p-values and the number of calibrating GCs 
used in the respective relations.}
\label{f:residual}
\end{figure*}

By looking at the residuals IQR1-IQR[O/Na] it is easy to understand what this
factor may be. Positive residuals are found for GCs mostly with a red HB
morphology (e.g. 47~Tuc, NGC~362, M~71), whereas negative residuals are more
common for GCs with a bluer HB (e.g. NGC~288, M~79, NGC~4833, M~80, M~12 etc.).
This fact gives us a hint to improve the calibration: the missing
factor must be related to the distribution of stars along the HB.

This conclusion is further strengthened by Fig.~\ref{f:residual} where the
residuals of the calibrated values IQR1 with respect to the observed IQR[O/Na]
are plotted as a function of various recent and older parameters describing the
HB morphology. The widely used HBR index measures the mean location in colour 
of the bulk of HB population through the ratio (B-R)/(B+V+R) where B, V, and R are
the numbers of blue, variable, and red HB stars. Recio-Blanco et al. (2006)
used the maximum temperature reached by stars on the bluest part of the HB,
$\log T_{eff}^{max}$, estimated with the zero-age HB models by Cassisi et al.
(1999). Dotter et al. (2010) computed the $\Delta (V-I)$ parameter as the median
colour difference between HB and RGB at the level of the HB in the GCs of the ACS
survey (Sarajedini et al. 2007). Finally, in the rightmost panel, the residuals
are plotted as a function of the integrated, reddening corrected, UV colours 
from the FUV e NUV bands observed with Galex (Dalessandro et al. 2012).

All the shown relations are statistically significant, some at a confidence
level exceeding 99.9\%, even when the parameters are available for a limited
number of calibrating GCs. It is clear that, beside structural  parameters such as
$M_V$ and $c$, the morphology of the HB is a parameter that must be included in
a calibration that aims to correctly reproduce the observed extension of the
Na-O anti-correlation.

\subsection{Final calibration}
There has been ample discussion on the best parameter for describing  the
distribution of GC stars on the HB, from the classical comparison of old and new
parameters by Fusi Pecci et al. (1993) to  more recent works introducing
variations or new parameters (Dotter et al. 2010,  Dalessandro et al. 2012,
Milone et al. 2014). 

Each operative definition comes with its pros and cons. Indexes tailored
to better represent the coolest part of the HB are affected by uncertainties
concerning the reddening estimates, blurring of AGB and RGB stars, possible
difficulties to individuate the terminal points of the red HB (see in particular
the discussion in Fusi Pecci et al. 1993, their Section 3.3). On the other hand,
other parameters are somewhat blind to the red extremes, being purposedly
defined to handle the hotter regions of the HB distribution. Among the
drawbacks, uncertainties in the end point of long blue tails, also related to
possible contamination by field stars. Parameters such as $L_{tail}$ (Fusi Pecci 
et al. 1993), $\log T_{eff}^{max}$ (Recio-Blanco et al. 2006), and L2 (Milone et
al. 2014) allow a finer HB-type resolution among GCs with long blue tails on the
HB. A common shortcoming is however that in most analyses they are measured on
optical (and sometimes old and shallow) CMDs, where the completeness of faint
extreme blue tail stars may become a relevant issue, as discussed by Dalessandro
et al. (2012).

After scrutinizing various options, we adopted the classical HBR index
for several reasons. This parameter may not be ideally suited for all HB types,
but is straightforward to compute, it is available for the vast majority of
Galactic GCs, and it is model independent. HBR values are also given for very small
GCs, usually neglected in most photometric surveys, allowing us to apply 
the calibration down to the very low mass regime. The index, or some close
variants, can be computed from deep HST CMDs of clusters even in the M31 galaxy
(see e.g. Perina et al. 2012), as well as in closer external galaxies such as the
Magellanic Clouds and the Fornax dSph (e.g. Mackey and Gilmore 2004). Finally,
the entry for HBR is also available for M~54, the second most massive GC in the
Milky Way, while no data are found for this cluster concerning other
parameters (except for L2, Milone et al. 2014). As well as sampling the high mass 
end of the GCs mass function, this provides the side benefit of including in the
calibrating dataset a cluster of extra-Galactic origin, M~54 being a confirmed
member of Sgr (Bellazzini et al. 2008). We may use this occurrence as a test for
the applicability of the results to other GC systems. Adopting this HB index 
we `lose' NGC~5139 ($\omega$ Cen) for which HBR is not given, but this is a
cluster with many peculiarities, besides hosting multiple populations (see e.g.
the  review by Gratton et al. 2004).

By adding the term in HBR, our finally adopted calibration is:

\begin{eqnarray}
IQR2 = -0.194(\pm 0.041) M_V -0.370(\pm 0.131) c \nonumber \\
+0.117(\pm 0.040) HBR -0.338(\pm 0.030)
\end{eqnarray}

with $rms=0.141$ and  $r_p=0.79$ (22 GCs). The improvement in this calibration 
with respect to IQR1 (eq. 1) is immediately evident: the two-tail probability 
to be a random result is now down to $p=1.2\times 10^{-5}$, the scatter is 
decreased, and the calibrated IQR2 explains $\sim 18\%$
more of the variability in the response (i.e. the extent of the Na-O
anti-correlation) due to adding the HB morphology description to the predictors.

\begin{figure}
\centering
\includegraphics[scale=0.40]{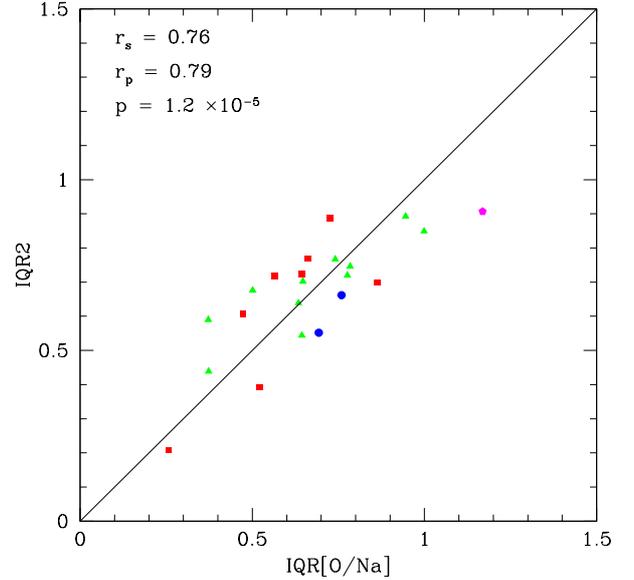}
\caption{Comparison of the IQR2 values from our final calibration  (equation 2)
with the observed values IQR[O/Na] for 22 calibrationg GCs. Symbols and labels
are as in Fig.~\ref{f:cfriqr1}.}
\label{f:cfriqr2}
\end{figure}

In Fig~\ref{f:cfriqr2} we compare the new calibrated values IQR2 to the observed
ones IQR[O/Na] for our 22 GCs. Now all the values satisfactorily lie around the
identity line: no statistically significant correlation of the residuals
IQR2-IQR[O/Na] as a function of global parameters is found. 

At first glance comparing Figs.~\ref{f:cfriqr1} and
\ref{f:cfriqr2}, we see that to correctly reproduce the one-to-one 
correspondence with the observed spreads of Na and O abundances it is necessary 
to introduce a correcting factor describing the distribution of HB stars in GCs.

On a second level, the goodness of this calibration bears a deeper physical
meaning. All the observed IQR[O/Na] values are computed in our calibrating
clusters from samples of red giants. The link between the chemical abundances of
light elements in RGB stars and the distribution in colour of stars in the next
evolutionary phase on the HB once again points towards the well known connection
existing between the multiple population phenomenon and the HB in GCs. 
Besides the structural parameters of a GCs there should be another ingredient
that is able to explain the coupling of different observations in two distinct
evolutionary phases, likely related to how multiple stellar populations formed
and evolved in the clusters.

This ingredient was predicted theoretically some time ago (D'Antona et al.
2002) and is already well known from monovariate relations. Carretta et al.
(2007a) find a very strong correlation between chemical variations in
proton-capture elements and the highest temperature reached on the ZAHB by GC
stars. Ever since, this relation holds for any new GC studied to date (see e.g.
Carretta and Bragaglia 2018). Enhanced He abundance was individuated as the
culprit: associated to stars with highly depleted O and enhanced Na abundances,
it results in decreased envelope masses, forcing the stars to spend the
following HB phase segregated at very high temperature regimes. If the
alterations in O and Na are due to proton-capture reactions in H burning at high
temperature, whose main outcome is of course He, then this is essentially the
relation connecting the variegated HB morphology and the multiple population
phenomenon in GCs.

\section{Application to Galactic GCs}

The calibration for IQR2 can be then applied to all Galactic GCs for which the
parameters $M_V$, $c$, and HBR are known. We considered the GCs listed in the
H10 catalogue with parameter $c <2.5$.
The HBR values were taken from Mackey and van den Bergh (2005), except for
Terzan~8, whose HBR is from Carretta et al. (2010a). For the division of GCs
into different Galactic sub-populations we followed the classification scheme
described in the Appendix A of Carretta et al. (2010a).

Excluding GCs without an HBR value, like NGC~5139 ($\omega$ Cen), and AM-1,  a
poorly studied cluster whose IQR2 turned out to be slightly  negative (probably
due to some problems in the parameters), we could apply Equation (2) and
provide new empirical predictions for the extent of the Na-O anti-correlation for
95 GCs. We also add the 22 GCs used as calibrators. Three more GCs in our FLAMES
survey (NGC~6397, NGC~6752, and NGC~7099=M~30)  are classified core-collapse;
while they were not used as calibrators, they have observed IQR[O/Na] values
directly derived by our group (Carretta et al. 2007c, 2009a,b) and we included
them in the sample. A grandtotal of 120 galactic GCs ($\sim 80\%$ of all 
classical GCs in the Milky Way) with extension of the Na-O anti-correlation on
the scale defined by our very homogeneous survey is then available. IQR2 values
for individual  GCs are listed in the Appendix, Table A.1.

\subsection{Sanity checks: Multiple populations, He, and HB}
\begin{figure*}
\centering
\includegraphics[scale=0.40]{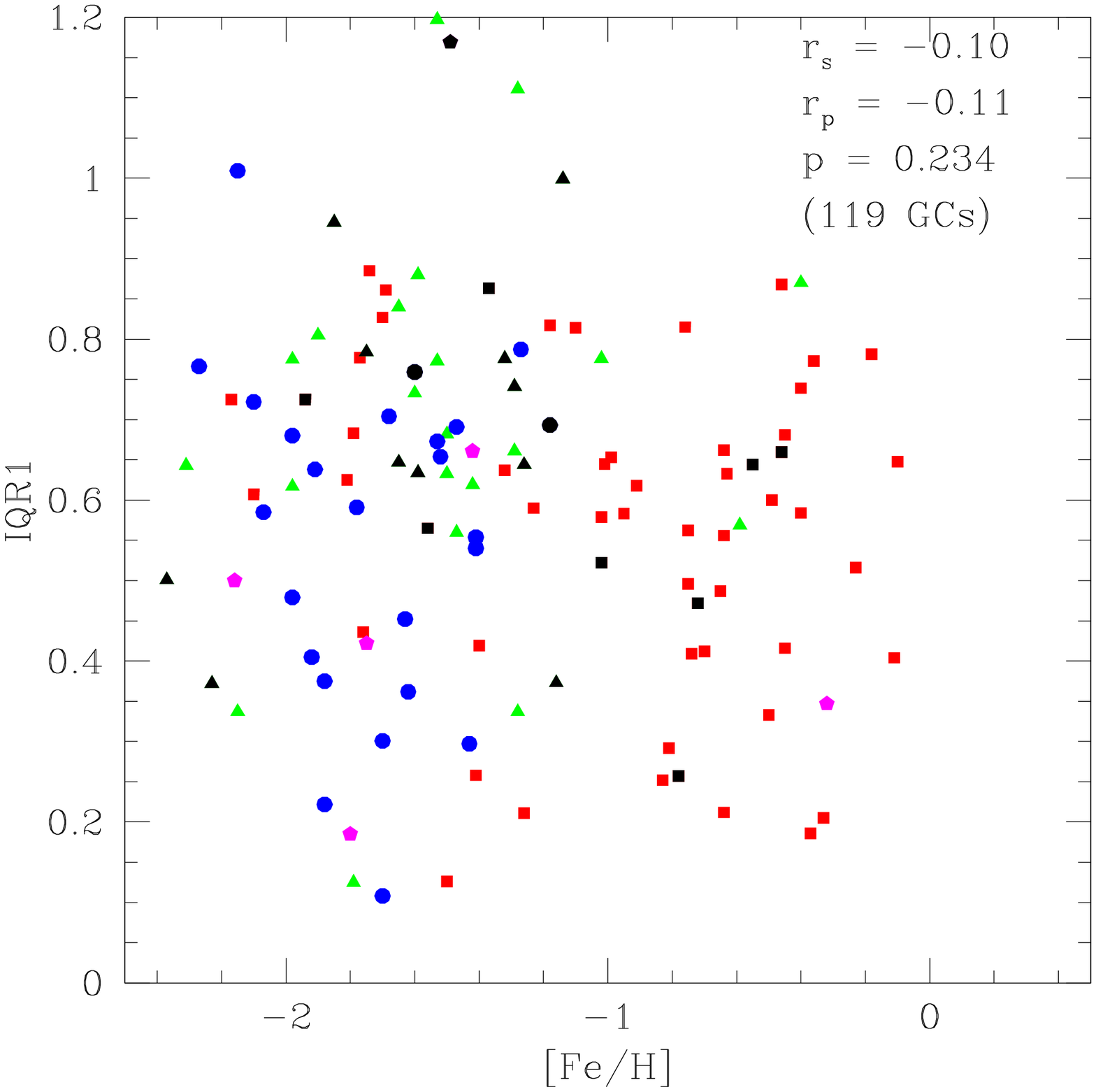}\includegraphics[scale=0.40]{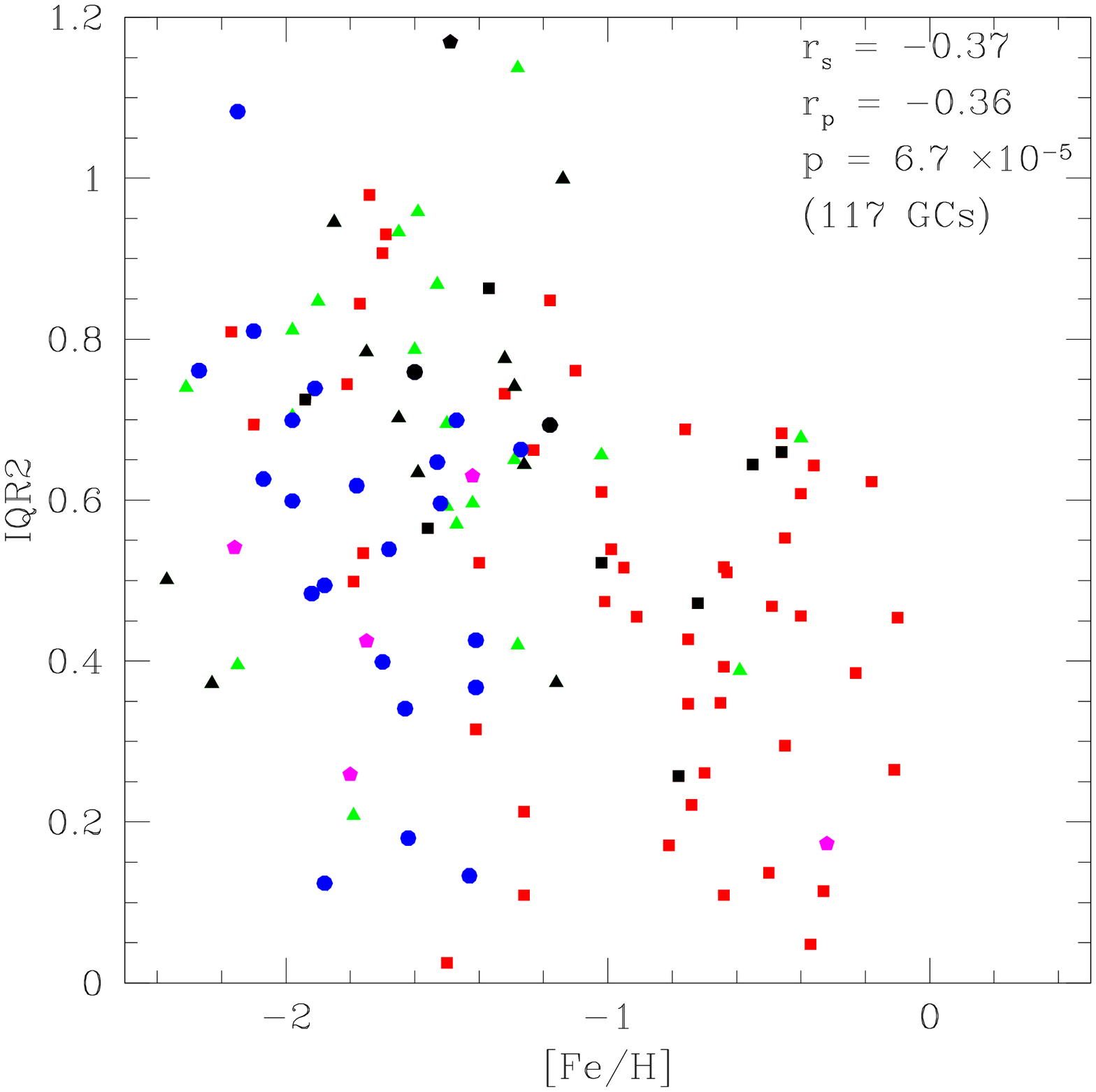}
\caption{Extensions of the Na-O anti-correlation IQR1 (left panel) and IQR2
(right panel) from the calibrations Eqs.(1) and (2), respectively. Symbols are
as in Fig.~\ref{f:cfriqr1}. Black points indicate the calibrating GCs from our
FLAMES survey, and their shape indicate the Galactic sub-population. 
Symbols and label are as in Fig.~\ref{f:cfriqr2}.}
\label{f:iqrfe}
\end{figure*}

This large dataset enables us to perform statistically robust tests 
to verify how reliable the final calibration is.
First, we compare in Fig.~\ref{f:iqrfe} the run of IQR1 (left panel) and IQR2
(right panel) as a function of the cluster metallicity [Fe/H]. As is evident in the
left panel, IQR1 does not show any correlation with metallicity. The Student's
t-test returns a two-tail probability of $p=0.23$. On the other hand, IQR2 is
found correlated with [Fe/H] and the relation is statistically very robust
($p=6.7 \times 10^{-5}$). 

All the main classes of polluters proposed to explain the spread of Na,
O (and other light elements) observed in GCs do not alter the cluster initial metallicity
(see Gratton et al. 2012, Bastian and Lardo 2018). Remarkably, even in the so-called iron
complex GCs, where intrinsic dispersions in [Fe/H] are measured, a Na-O
anti-correlation is detected in each individual metallicity  component 
(e.g. NGC~1851, Carretta et al. 2011; M~22, Marino et al. 2011; NGC~6273, 
Johnson et al. 2015). 
When the term for the HB enters in the calibration for IQR2, so does implicitly
also the dependence on metallicity, which is the first parameter driving the HB
morphology. This self-consistently explains the correlation (or the lack
thereof) observed in Fig.~\ref{f:iqrfe}.

A second consistency check for our calibrations is based on the He variations
expected in multiple stellar populations.  Stellar models He-enhanced reach the
H-exhaustion point earlier than their counterparts with lower He. For a mixture
of stellar  populations differing in their He content we must then expect a spread in colour on 
the main sequence (MS), or even split sequences in the most
extreme cases (e.g. Milone et al. 2012a,b). Thus, if our calibrated extents of
the Na-O anti-correlation correctly account for concomitant enhancement of He 
after the nuclear processing producing Na-enhanced, O-depleted matter, a
correlation with the colour spreads along the MS should be  seen.

\begin{figure*}
\centering
\includegraphics[scale=0.40]{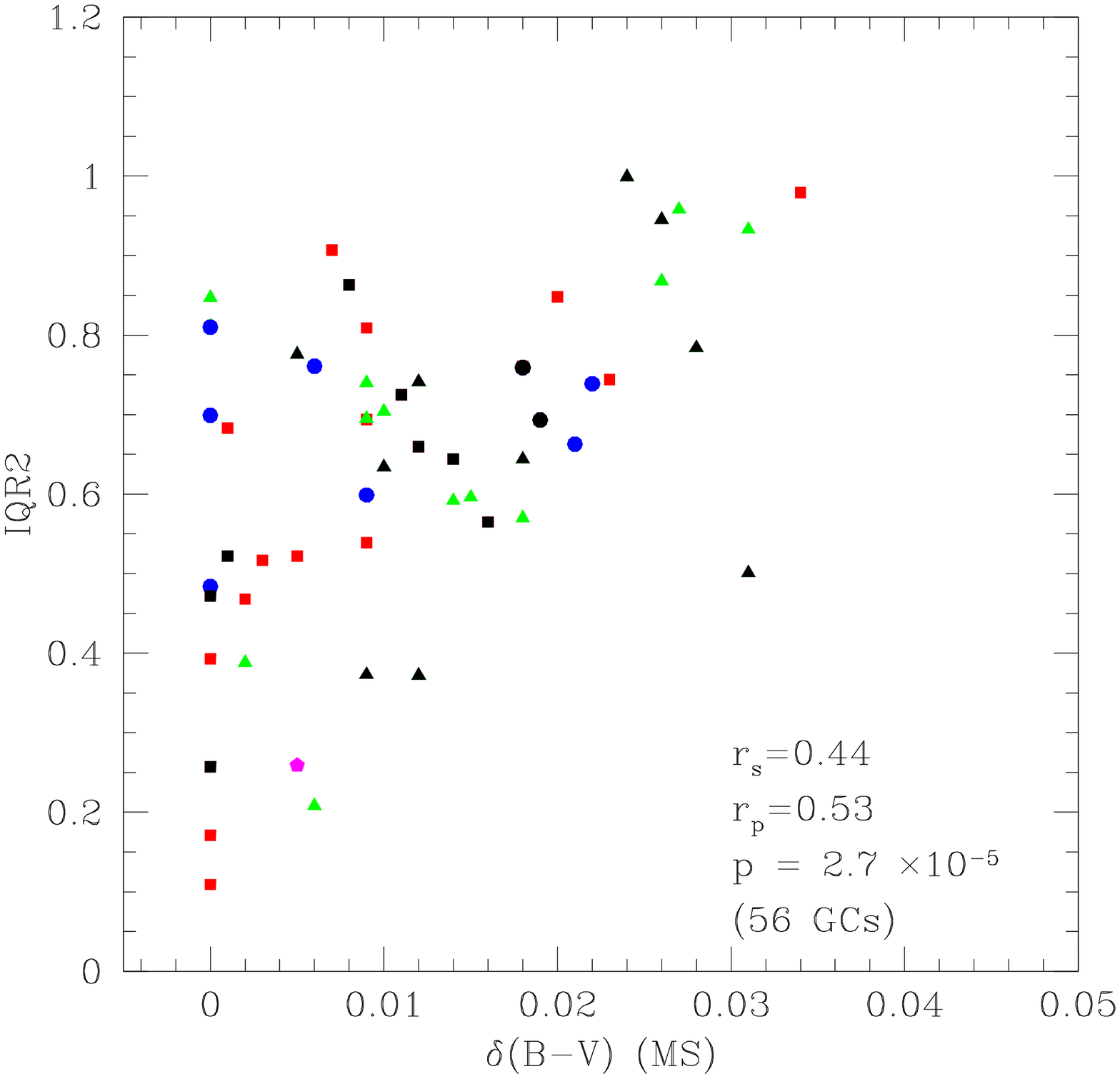}\includegraphics[scale=0.40]{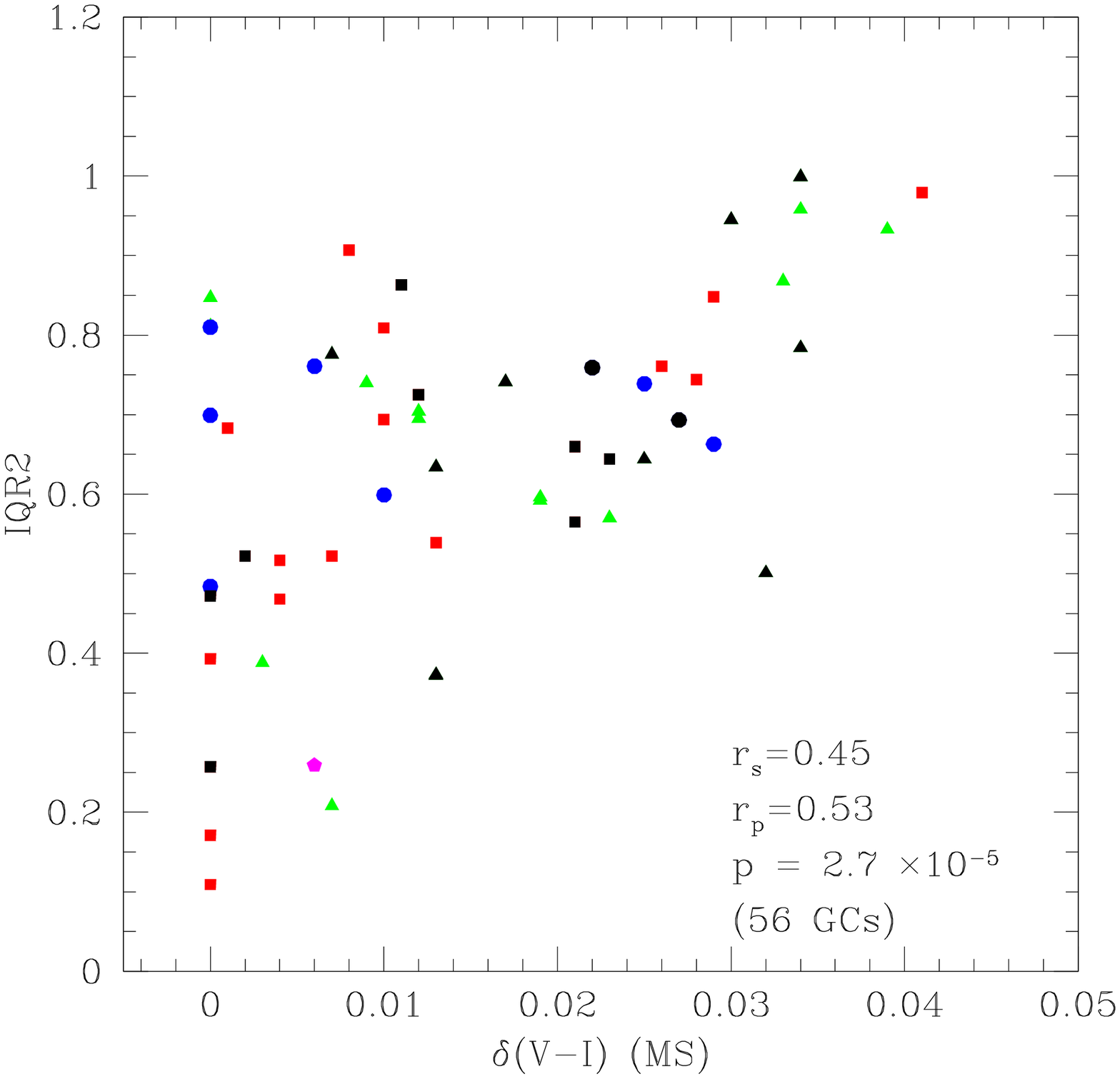}
\caption{Calibrated extent of the Na-O anti-correlation IQR2 as a function of
the colour spreads in $B-V$ and $V-I$ along the MS from Gratton et al. (2010).
Symbols and labels are as in the previous figures.}
\label{f:iqrMS}
\end{figure*}

For 56 GCs in our dataset Gratton et al. (2010) estimated the colour spreads 
$\delta (B-V)$ and  $\delta (V-I)$ expected along the MS from the computed He
spreads corresponding to the mass dispersion along the HB. These two parameters,
derived independently, are found to be strongly correlated with our  empirical
estimates of the Na-O anti-correlation extent in Fig.~\ref{f:iqrMS}. Again,
different approaches (spectroscopic and photometric) at two different
evolutionary phases (RGB and MS) concur to the same general, consistent picture.

As a third consistency check, we can test whether a correlation is found with more direct estimates of
He abundances. However, the limitations for a spectroscopic derivation of the He
content in GC stars are even more severe than those concerning O and Na, both
among HB and RGB stars (see discussions in Gratton et al. 2010, Dupree et al.
2011, Pasquini et al. 2011,  Milone et al 2018) so that it is still necessary to
resort to vicarious measurements. Recently, Milone et al. (2018) have produced
estimates of the maximum He variation in 57 GCs by comparing multi wavelength
$HST$ photometry to synthetic spectra  computed with abundances appropriate for
FG and SG stars. 

\begin{figure}
\centering
\includegraphics[scale=0.40]{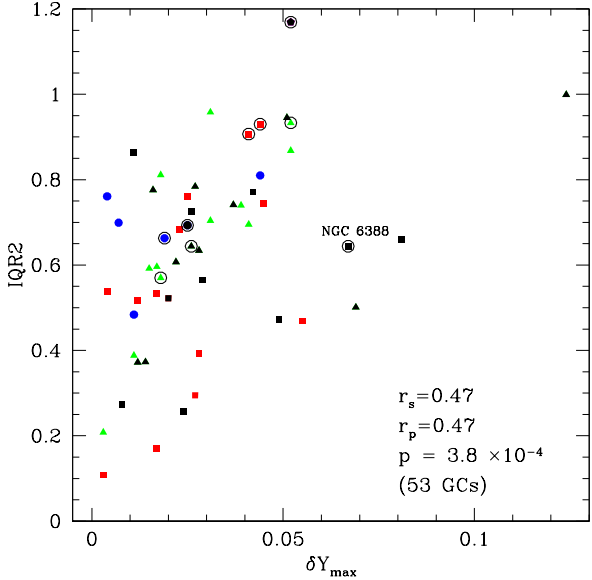}
\caption{Calibrated IQR2 values for the extension of the Na-O anti-correlation as
a function of the maximum variations in He derived from photometry (Milone et 
al. 2018). GCs defined as type-II are encircled, and the position of NGC~6388 is
labelled (see text).}
\label{f:iqr2dymax}
\end{figure}

In Fig.~\ref{f:iqr2dymax} our calibrated values IQR2 are plotted against their
$\delta Y_{max}$ estimates for 53 GCs in common. There is a strong correlation,
significant at a level of confidence exceeding 99.9\%. There is also a
noticeable scatter, and for this we offer two explanations.
First, as already discussed in Fusi Pecci et al. (1993), the effects of a
secondary parameter may result in correlations that appear to be less clean
than those generated by primary parameters. The dependence of IQR2 on He is
mediated by the HB morphology, because the ZAHB evolution of He enhanced stars
is preferentially spent at bluer locations. However, the He content was tagged
only as a third parameter for the HB by Gratton et al. (2010).

Second, beside the statistical impact, there could be a more physical meaning,
because of the different dependence on stellar mass for the alterations and
following release of He, O, and Na in the polluting matter (see e.g. Gratton 
et al. 2010, 2012). Let us consider the two most favourite candidate polluters,
intermediate-mass AGB stars (Ventura et al. 2001) and fast rotating massive
stars (FRMS, Decressin et al. 2007). For both candidates the phase at which He is
produced is the same: the main sequence. However, the production and release of
He is accompanied by O-depletion and Na-enhancement at the same evolutionary
phase in FSRM, while it is delayed until the hot-bottom burning in the AGB
phase in the AGB scenario, where the production of He is somewhat decoupled
from that of O and Na.
Thus, the trend and scatter in Fig.~\ref{f:iqr2dymax} could perhaps be seen as
the action of AGB polluters or possibly even of a mixture of polluters, with
FRMS working at the very early phases and the contribution from AGB stars 
entering at the appropriate (longer) timescale. We note that the estimates
$\delta Y_{max}$ from Milone et al. (2019) show significant correlations 
with the colour spreads $\delta (B-V)$ and  $\delta (V-I)$ from Gratton et al.
(2010: two-tail p=0.0022 and 0.0010, respectively), although with some scatter.

Finally, we note that the GCs in Fig.~\ref{f:iqr2dymax} seem to 
approximatively follow two different branches, and eight out of nine type-II GCs (as
defined in Milone et al. 2017, and marked with open circles in our figure) lie
on the upper branch. These are the GCs usually called iron-complex or anomalous
from their intrinsic dispersion in Fe, as well as heavier elements like those
from s-process. Photometrically, almost all type-II  GCs show split sub-giant
branches (SGBs), with the faint SGB continuing up to a red RGB. As mentioned
above,  in these GCs the multiple population phenomenon appears in each
metallicity component (e.g. Carretta et al. 2011, 2010b,c, Marino et al. 2015,
Johnson et al. 2015 for NGC~1815,  M~54, NGC~5286, and NGC~6273, respectively).
The only exception in Fig.~\ref{f:iqr2dymax} is NGC~6388, clearly located on the
lower `branch'. However, from high resolution spectroscopy there is no sign of
an intrinsic dispersion in [Fe/H] in this 
massive GC ($\sigma_{\rm [Fe/H]}= 0.041$ dex from the direct analysis of 24 RGB stars 
in Carretta and Bragaglia
2018 and $\sigma_{\rm [Fe/H]}= 0.051$ from 151 giants in Carretta and Bragaglia
2019, in prep.; both values are fully compatible with observational errors only). In
passing we also note that NGC~6388 is the only GC defined as type-II for which
Milone et al. (2017) could not clearly establish the connection faint SGB-RGB. 

Hence, it seems that all genuine iron complex (or type-II) GCs are
segregated on the (tentative) upper branch in Fig.~\ref{f:iqr2dymax}. Taken at
face value, this observation indicates that these GCs have much-too-high IQR values if compared to their estimated maximum variation in He. 
While a deeper study of the full links between spectroscopic and photometric
properties of GCs is desirable, it is beyond the immediate purposes of the
present work. We limit the extent of our study here, and summarize the results of this section below.

The extension of the Na-O anti-correlation can be calibrated as a function of
structural parameters of the GCs. The observed values of spreads in O and Na are
best reproduced when the HB morphology is accounted for in the empiric
calibration. Its reliability is corroborated because the calibration correctly returns a
dependence on metallicity, the first parameter governing the HB morphology.
Moreover, the He enhancement predicted to go along with the alterations in O and Na
is well incorporated by the calibration, as shown by the correlation of IQR2
with the colour spreads on the MS phase and with estimates of He variations
deduced from UV photometry. 

We suggest that the spread observed in the relation between the extent of the 
Na-O anti-correlation and the He spread in GCs may be ascribed either to the
decoupling in AGB stars between the peaks of He and O, Na changes in the
polluting matter or to the presence of different classes of polluters acting at
different times. The last hypothesis is endorsed by several existing cases of 
discrete groups observed along the anti-correlations of proton-capture
elements, where the stellar components with extreme and intermediate composition
cannot be reproduced using only one class of massive polluters (NGC~6752:
Carretta et  al. 2012; NGC~2808: Carretta et al. 2018; NGC~6388: Carretta and
Bragaglia 2018).

All these tests confirm the coherent picture of multiple stellar populations in
GCs, where O-depleted and Na, He-enhanced stars help to simultaneously explain
the observed extent of the Na-O anti-correlation, the distribution of stars in HB
and the spreads in colour along the MS. We can thus be confident that the 
adopted empirical calibration returns a reliable, quantitative estimate of the
nuclear processing and chemical modifications occurred at early times in GCs.

\subsection{Na-O anti-correlation across GCs in the Milky Way}

With the above calibration, empirical estimates for the Na-O anti-correlation
can be derived for 120 Galactic GCs, including the most distant or smallest 
ones, that up to now were not investigated for multiple stellar populations 
neither with spectroscopic nor photometric means.

\begin{figure}
\centering
\includegraphics[scale=0.40]{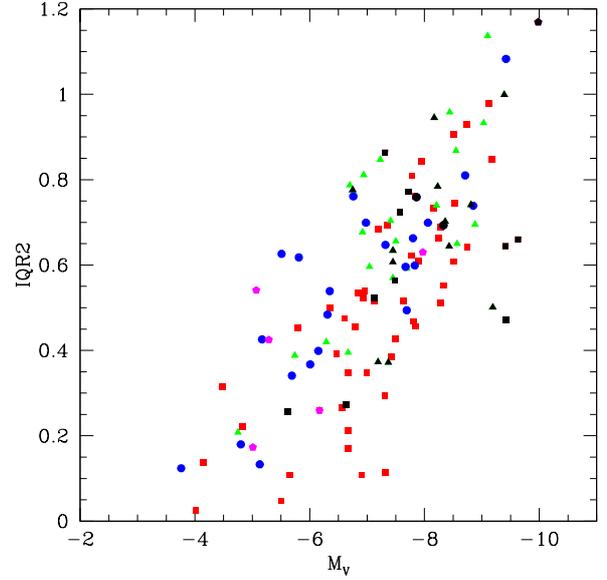}
\caption{Calibrated IQR2 values from Eq.(2) as a function of the total absolute
magnitude from H10. Observed IQR[O/Na] values are used for 25 GCs from our
FLAMES survey (black symbols).}
\label{f:iqr2mv}
\end{figure}

In Fig.~\ref{f:iqr2mv} we plot the calibrated IQR2 values as a function of the
total absolute magnitude $M_V$ from H10. For 25 GCs analysed in our FLAMES
survey in this figure we use the observed IQR[O/Na] values (black symbols).

The high-luminosity end of our relation is secured by the presence of  the
massive GC M~54 in the calibrating sample. Formally, our empirical calibration
(Equation 2) should be valid only down to the faintest $M_V$ magnitude in our
calibrating sample, corresponding to NGC~6838=M~71 (black square at $M_V =
-5.61$). Hence, for about a dozen GCs we are formally in an extrapolation
regime. We note, however, that no indication of departures from linearity is evident
from both Fig.~\ref{f:cfriqr2} and  Fig.~\ref{f:iqr2mv}. The increased scatter
in the latter may be well ascribed to increasing uncertainties in the literature
estimates of parameters. Supporting evidence is given by two GCs below the above
threshold, Pal~5 and Terzan~8  ($M_V = -5.17$ and $-5.07$, respectively). In
both these small clusters SG stars are directly found using high resolution
spectroscopy (Smith et al. 2002 and Carretta et al. 2014b, respectively),
providing an unambigous chemical tagging, although for very few stars. 

With these caveats in mind, we can provide estimates of IQR2 for GCs as faint as
Pal~13 ($M_V = -3.76$). Formally, for any spatially resolved GC whose structure
can be reasonably approximated by King (1966) models it is possible to derive a
value for the  extension of the putative Na-O anti-correlation due to hosted
multiple populations.  Due to the dependence on different parameters in Eq.(2),
however, the smallest extent we estimate does not correspond to Pal~13. The
record holder seems to be the GC 1636-283, also known as ESO452-SC11, with IQR2=0.025.
This tiny bulge GC (estimated mass of only a few thousand solar masses,
Simpson et al. 2017) lies at the mass border between globular  and open
clusters, so the small derived value for IQR2 could simply be an artefact of the
calibration, mistaken for a real value. Fortunately, the recent spectroscopic
analysis by Simpson et al. (2017) shows star-to-star abundance variations in
this GC, not only from CN absorption features, but also from direct measurements
of Na abundances. Incidentally, this example well points out the strength of our
calibration: Simpson et al. complain that it is highly unlikely that many
other member stars in this GC can be reached for abundance analysis, whereas 
an estimate of inhomogeneities is easily obtained from our approach.

Among the GCs in Fig.~\ref{f:iqr2mv} with the smallest IQR2 values, another 
noteworthy cluster is Liller~1. With $M_V=-7.32$ and IQR2=0.114 this object is
located a little way off the main locus defined by the other GCs. However, a relevant
revision of its structural parameters has been proposed by Saracino et al.
(2015), who found Liller~1 to be less concentrated than previously estimated 
($c=1.74\pm 0.15$ instead of 2.3). From Eq.(2) this update would change the
value of IQR2 from 0.114 to 0.321, more in agreement with the moderately high 
mass estimated for this GC.

\begin{figure*}
\centering
\includegraphics[scale=0.40]{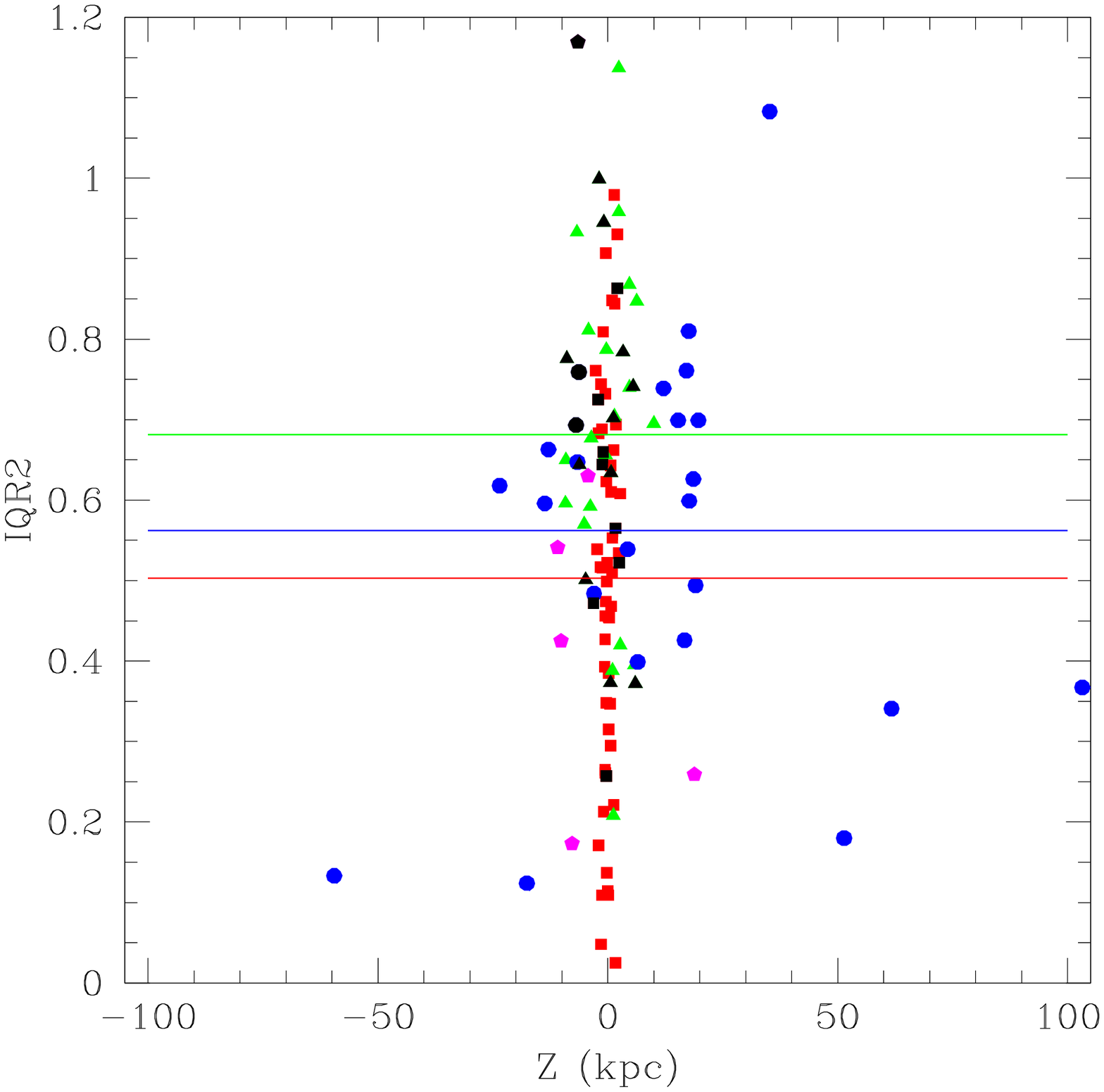}\includegraphics[scale=0.40]{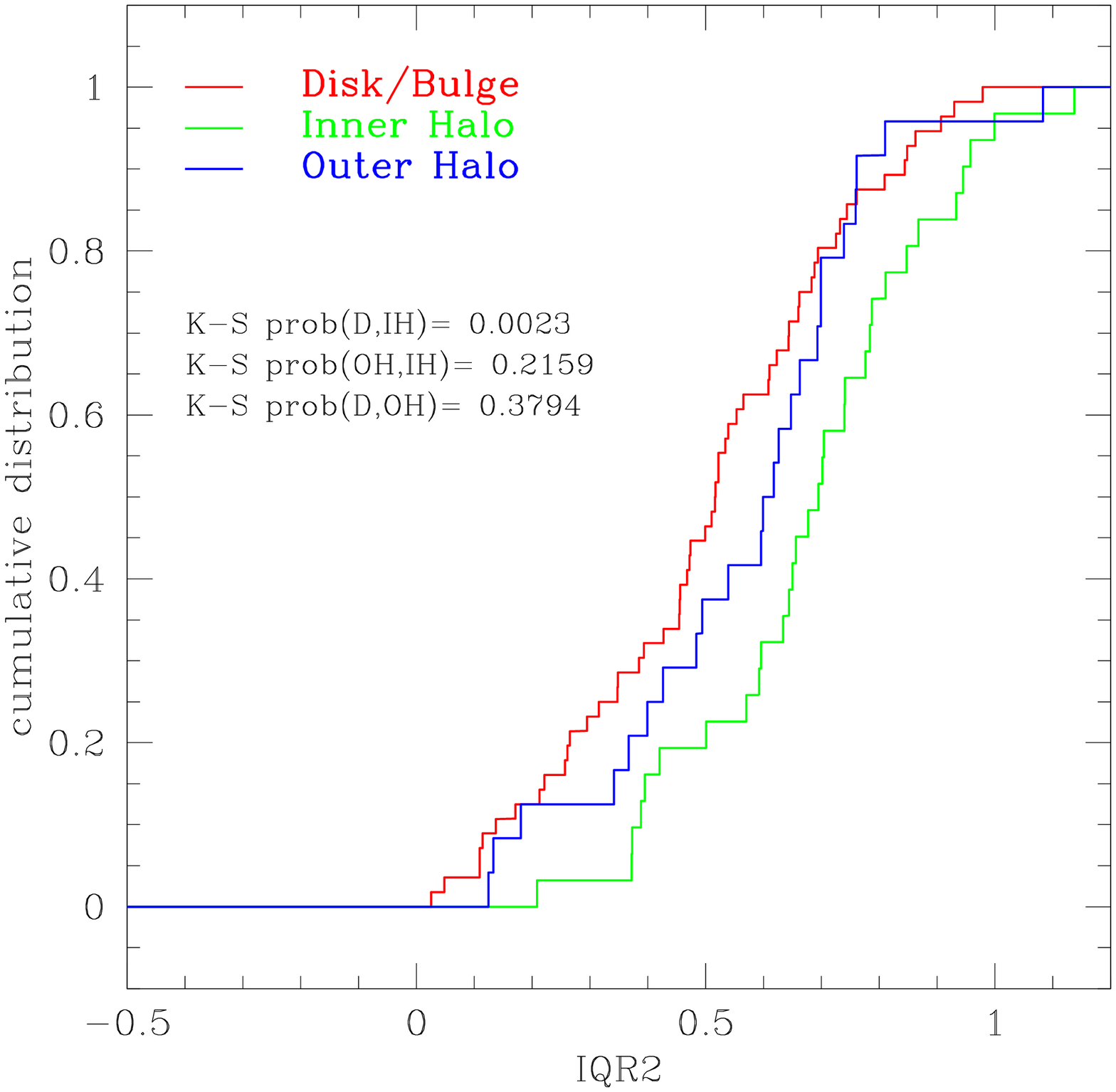}
\caption{Left: Calibrated IQR2 values as a function of the distance in kpc from
the Galactic plane (H10). Horizontal lines are traced at the average values of
the corresponding galactic  populations (same colour coding). Right: Cumulative
distribution of IQR2 for GCs of different sub-populations. The P values from
two-sample Kolmogorov-Smirnov (K-S) tests are indicated.}
\label{f:iqr2Z}
\end{figure*}

To compare the derived extension of Na-O anti-correlation in GCs of different 
Galactic sub-populations in Fig.~\ref{f:iqr2Z} (left panel) we plotted IQR2  as
a function of the the distance in kpc from the Galactic plane. We also show  the
average values for the individual populations of bulge/disc, inner, and outer
halo GCs (BD, IH, and OH): 0.503 ($rms=0.240$, 58 GCs), 0.678 ($rms=0.209$, 32
GCs), and 0.562 ($rms=0.227$, 24 GCs), respectively. A Student's $t-$test shows
that the only statistically significant difference is between the mean values 
for BD and IH GCs ($t=3.60$, 88 degrees of freedom, two-tailed probability
$P<0.001$), as also confirmed by a Kolmogorov-Smirnov test on the cumulative
distributions of IQR2 (Fig.~\ref{f:iqr2Z}, right panel).

\begin{figure*}
\centering
\includegraphics[scale=0.30]{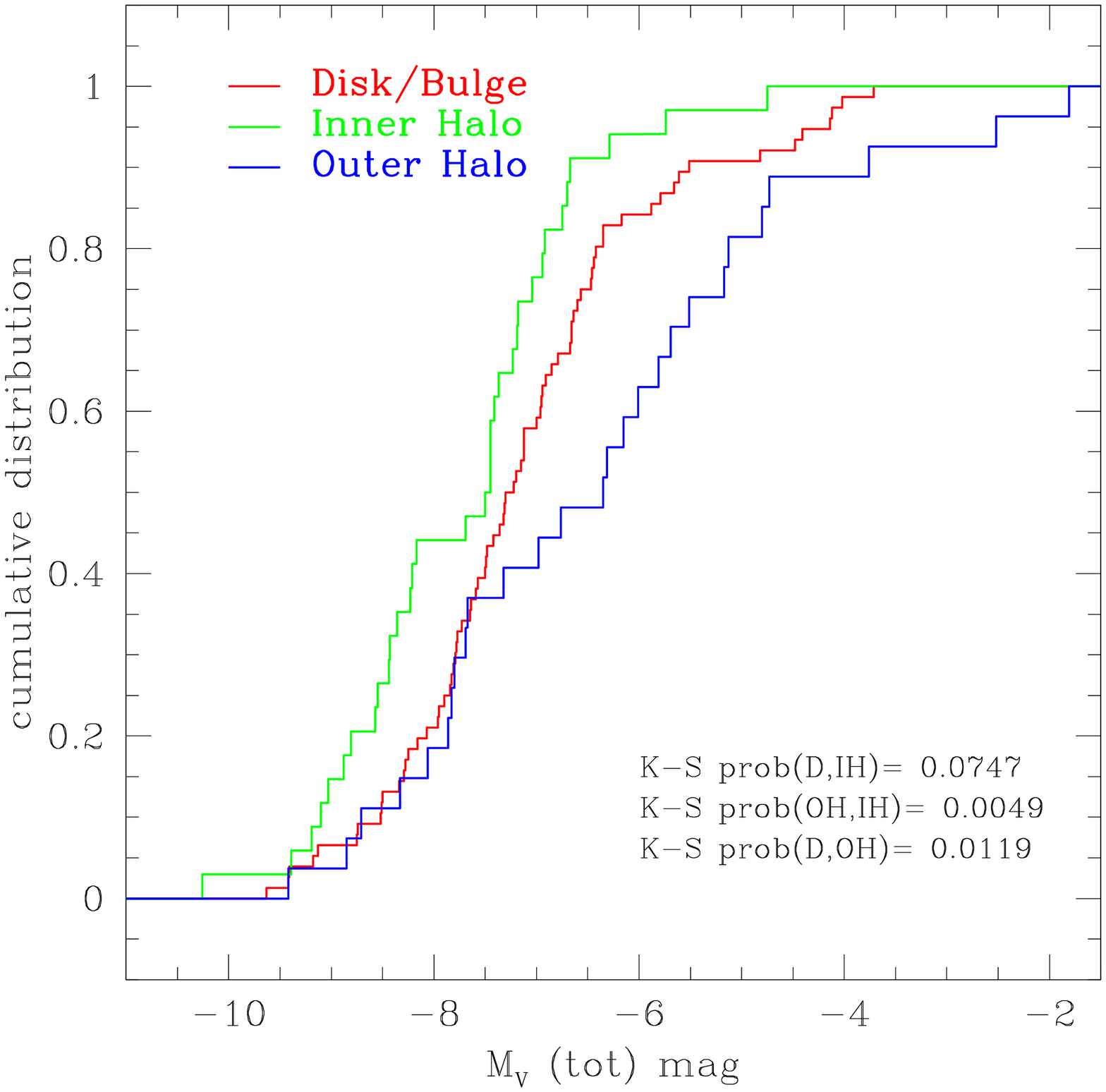}\includegraphics[scale=0.30]{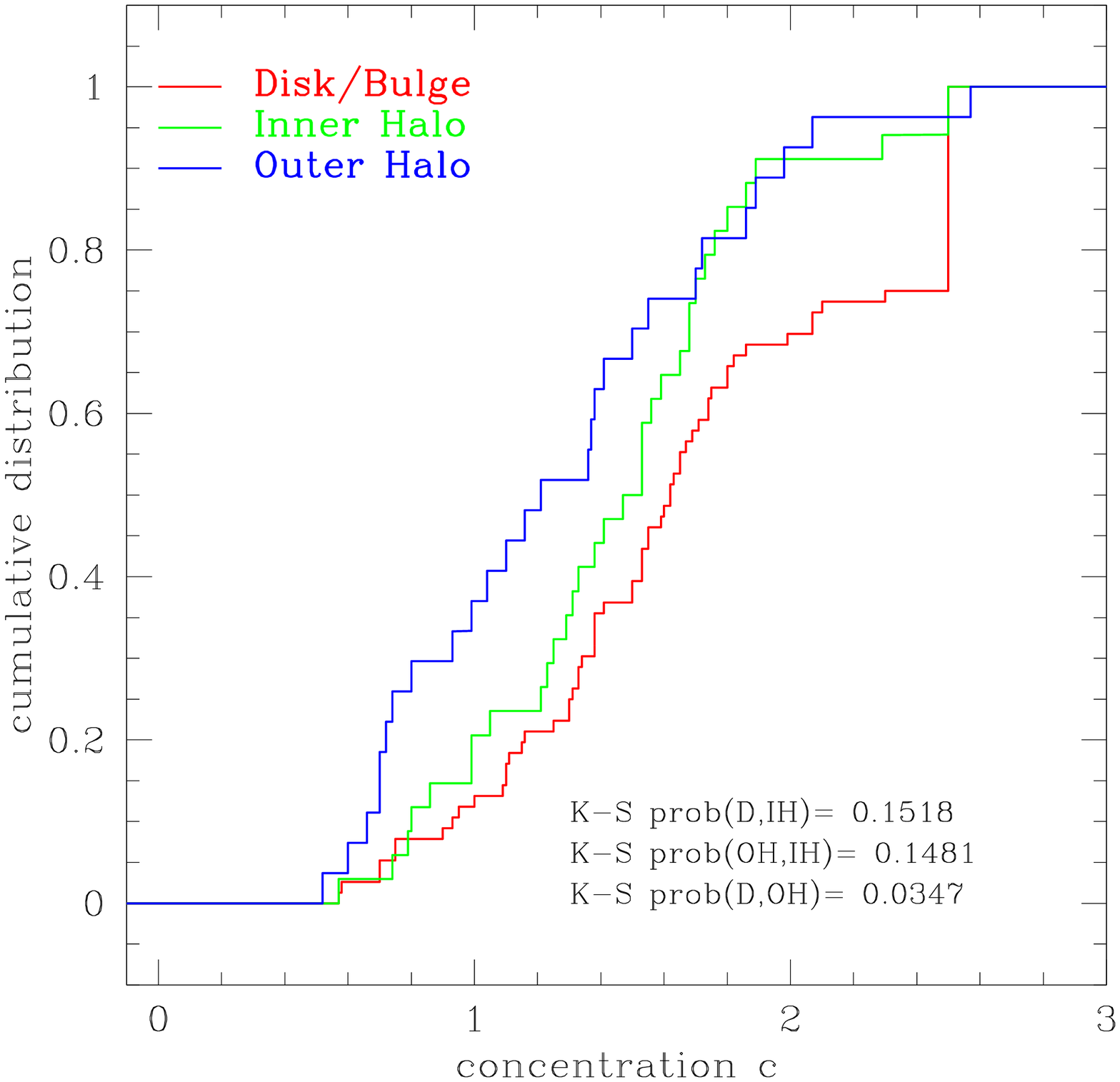}\includegraphics[scale=0.30]{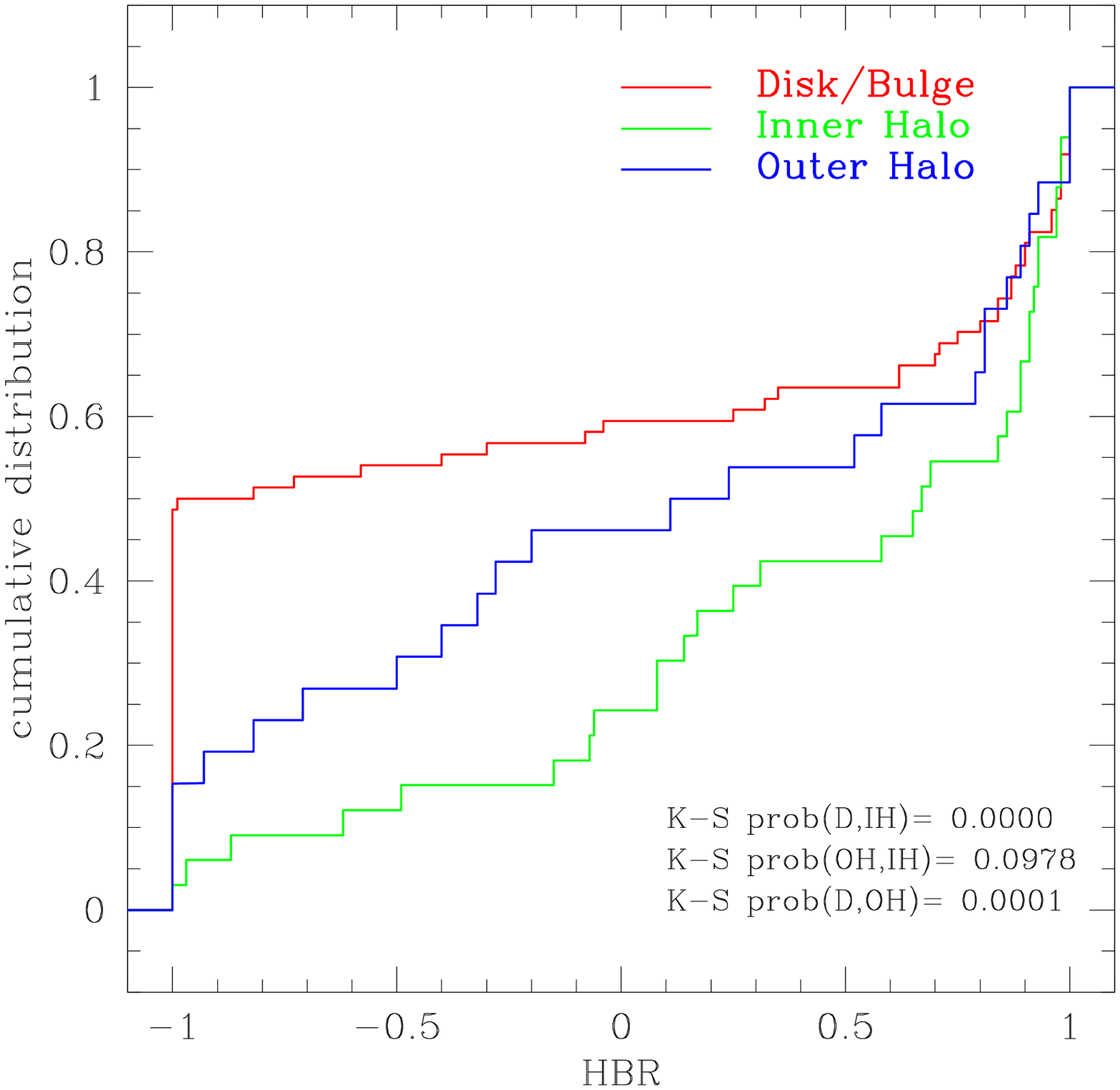}
\caption{Left, middle, and right panels: Cumulative distributions of total
absolute magnitude $M_V$, concentration $c$, and HBR index for GCs of different
Galactic sub-populations. Probabilities from K-S tests are indicated in each
panel.}
\label{f:cumupar}
\end{figure*}

However, before concluding that (inner) halo GCs did experience, on average, a 
greater degree of processing than bulge/disc GCs, as would seem judging from the
extent of the Na-O anti-correlation, we caution that it is necessary to look at
the intrinsic properties of the parent cluster sub-populations.

This is done in Fig.~\ref{f:cumupar}, where the cumulative distributions of the
parameters entering in the calibrating Eq. (2) are shown. From the luminosity
functions in the left panel, the inner halo hosts more luminous (i.e. massive)
GCs than both outer halo and disc. Many small GCs are found in  the disc,
contrary to the common sense of expecting small GCs more efficiently destroyed
in the more crowded and denser central galactic regions. As discussed  in, for example,
Carretta et al. (2010a), this evidence suggests that  differences in LFs might
be more related to the formation mechanisms than to destruction processes, 
although this may even not be the whole story. From the middle panel of
Fig~\ref{f:cumupar} we also may appreciate that GCs in the disc are more 
compact than those of both inner halo and outer halo, and are better suited to resist
to disruption in the ensuing evolution. Finally, disc GCs have basically redder
HB morphologies than outer halo GCs, whereas in turn IH GCs have bluer HBs.
These properties of IH GCs, in particular the coupling between
high GC mass and blue HB morphology, have been known for more than ten years (Recio-Blanco et al. 2006, Lee  et al. 2007).

We conclude that the larger IQR2 values found for IH GCs are fully consistent
with the distribution of parameters appropriate for this particular Galactic
sub-population. In Carretta et al. (2010a) from a limited sample of 19 GCs we
concluded that mass and metallicity are among the main parameters driving the
chemical signatures of multiple populations, with an additional contribution
given by the location in the Galaxy. The proposed empirical calibration and
the discussion above allow a better understanding of why this occurs: the extent
of the chemical processing between different stellar generations is modulated
by the distribution of structural parameters and HB morphology (whose first
parameter is indeed the metallicity).

\begin{figure*}
\centering
\includegraphics[scale=0.40]{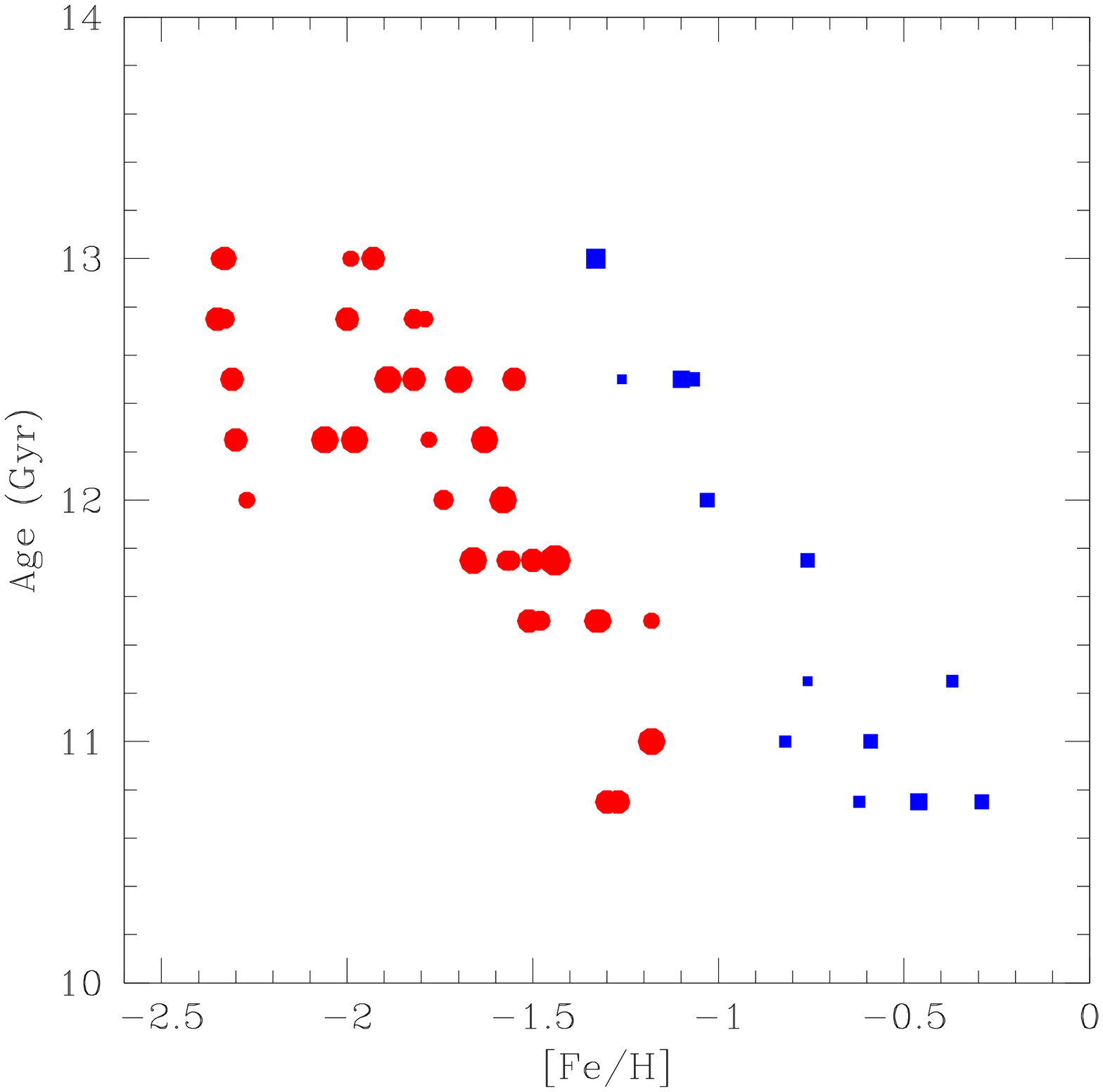}\includegraphics[scale=0.40]{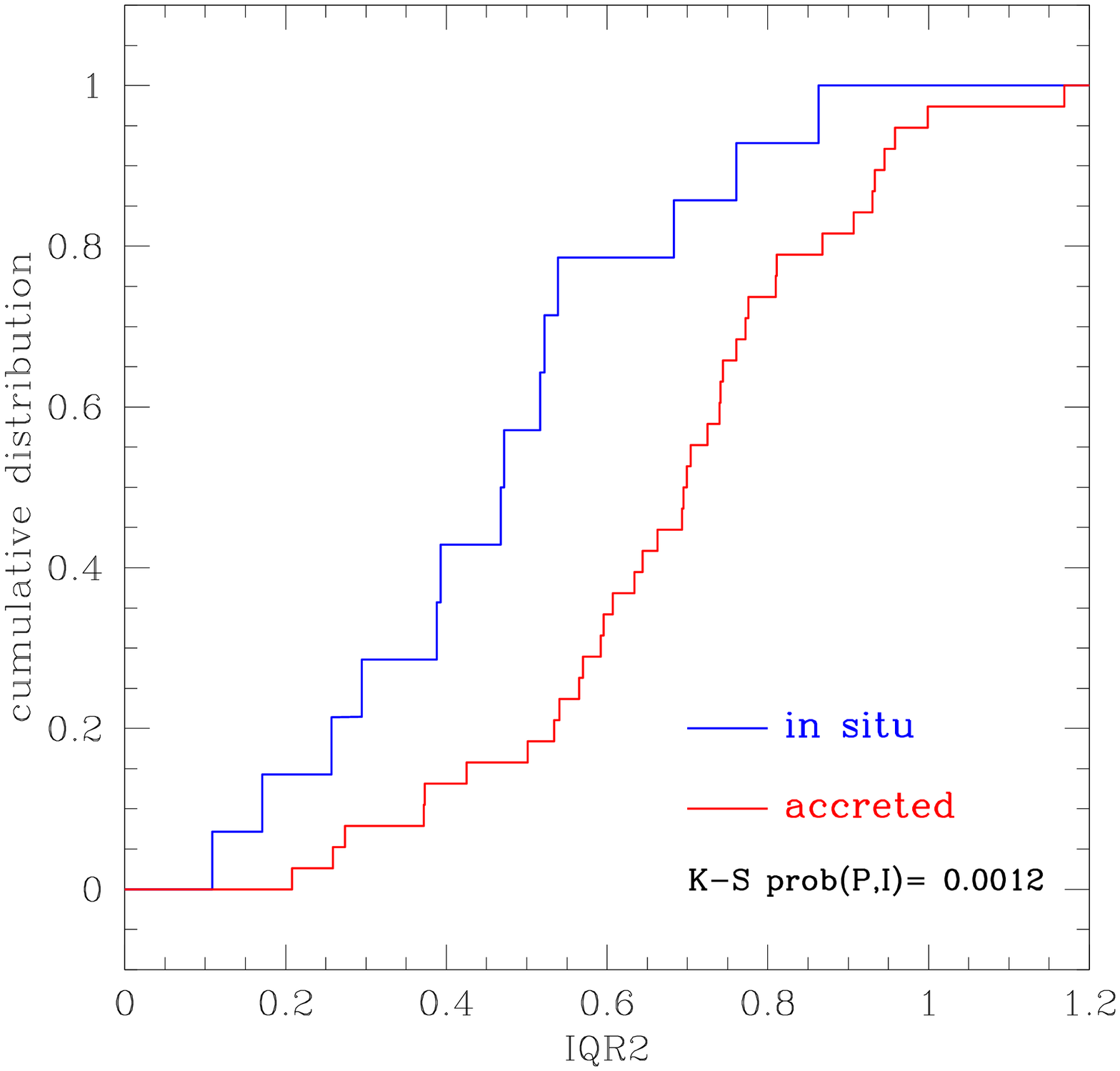}
\caption{Left: Age-metallicity relation from Leaman et al. (2013) for GCs formed
in situ in the MW disc (blue points) and GCs likely formed in  smaller stellar
systems and later accreted in the MW (red points). The size of the symbols are
proportional to the IQR2 values. Right: Cumulative distributions of GCs with
different origin. The P value from two-sample K-S tests is indicated.}
\label{f:accreted}
\end{figure*}

The same line of reasoning does apply to the difference in the estimates of the 
Na-O anti-correlation in GCs with a different origin. In Fig.~\ref{f:accreted}
left panel) we reproduced the age-metallicity relations obtained by Leaman et
al. (2013) for GCs formed in situ in the Galactic disc and those likely
originated in smaller systems, later accreted and disrupted in the Milky
Way. The size of the  symbols in this figure are proportional to the IQR2
values, and clearly indicate that the extension of the Na-O anti-correlation is
larger in the accreted component, as is confirmed statistically by the K-S test on
the cumulative distributions plotted in Fig~\ref{f:accreted} (right panel).
It is easy to verify, using the classification provided by Leaman et al., that the
accreted GCs are more massive and have bluer HBs than the GCs formed in situ.
Again, the differences in the extent of the multiple population phenomenon are
driven and explained by the mean properties of the parent cluster populations, 
even concerning the environment where they originally formed, as in this
example.

\section{Discussion}

The case of the globular cluster Pal~5 is emblematic to better understand the
dependence of IQR2 from structural cluster parameters. The extended tidal tails
associated to Pal~5 (e.g. Odenkirchen et al. 2001, 2003) are a direct evidence
of a large mass loss due to severe tidal disruption in the potential well of our
Galaxy. Clearly the present-day total mass of GCs (as portrayed by the total 
luminosity) is only a lower limit of the initial GC mass. Yet, not only direct
proofs of SG composition do exist for Pal~5, but also our calibration for IQR2
provides a value for the extension of the Na-O anti-correlation (IQR2=0.426) which
locate this GC right in the middle of the relation defined by all GCs.
We argue that our calibration represents a good empirical way to approach the
initial conditions of GCs, where the dependence on the (unknown) initial mass is
included in the present day mass and all the Gyr-long history of dynamical
evolution due to internal and external dynamical processes is modelled by the
concentration parameter.

\begin{figure*}
\centering
\includegraphics[scale=0.40]{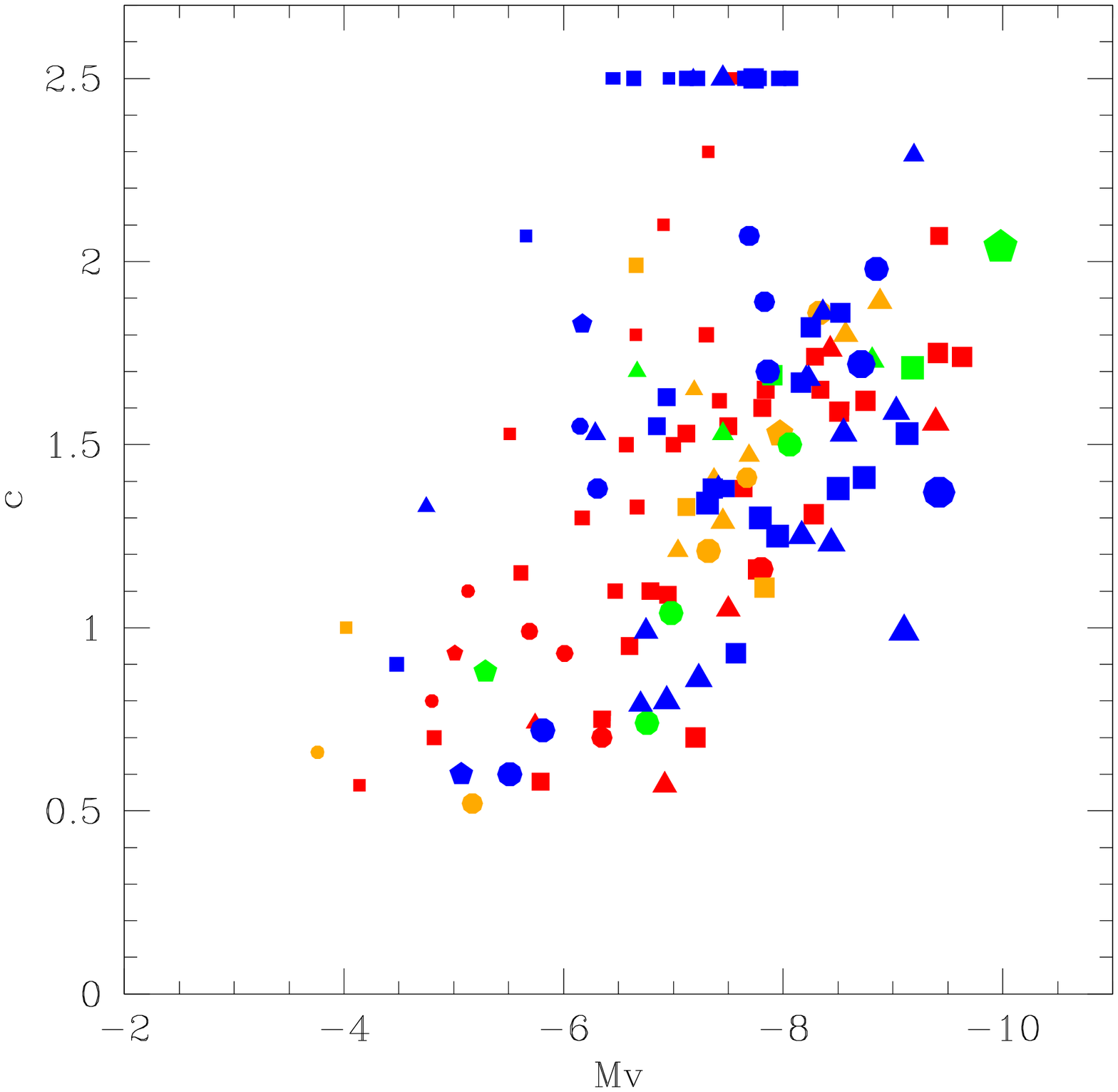}\includegraphics[scale=0.40]{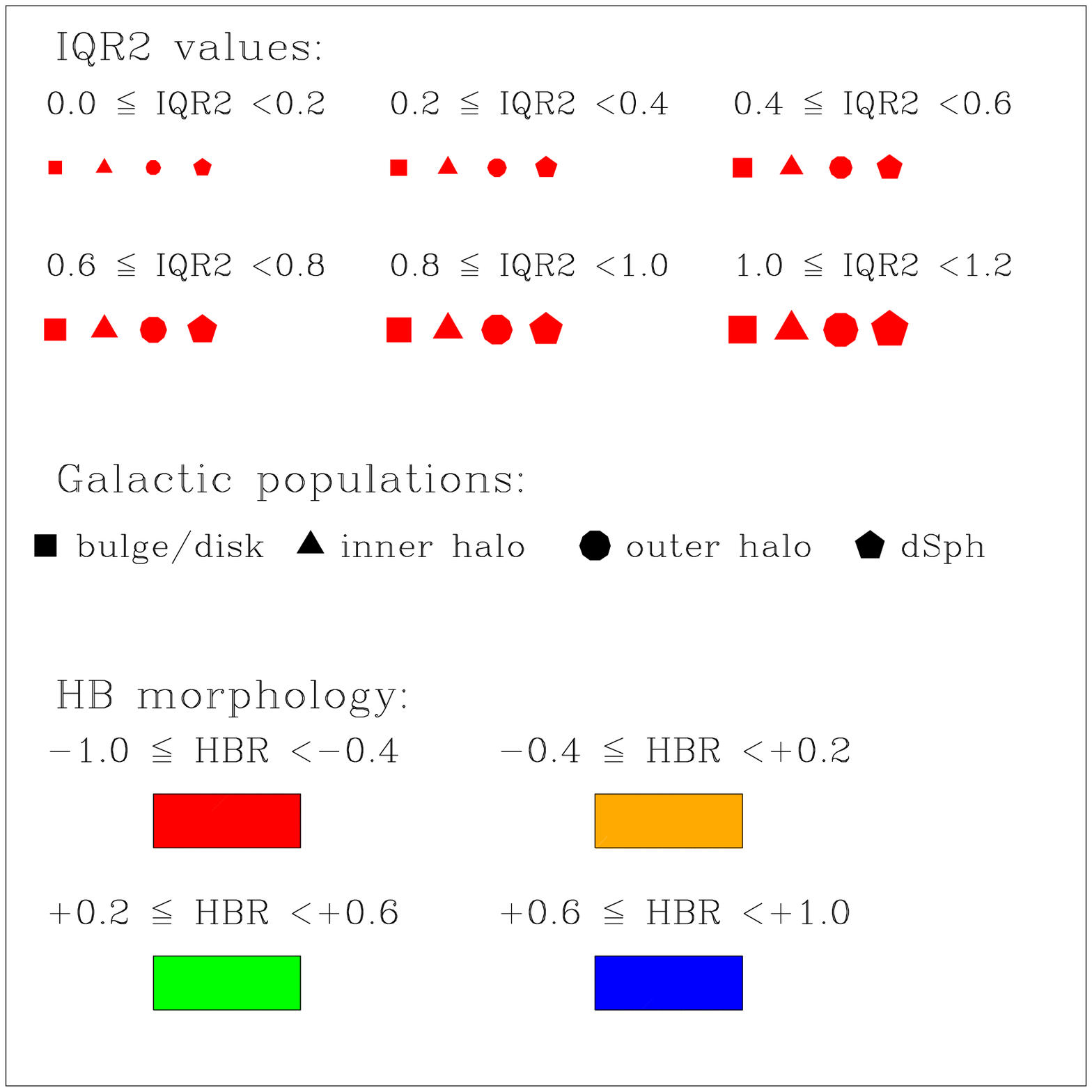}
\caption{Left: correlation between concentration $c$ and total absolute
magnitude $M_V$ for MW GCs. The size of symbols is proportional to the IQR2
value, their shape indicates the Galactic population, and the colours refer to 
the HB morphology, as detailed by the legend in the right panel.}
\label{f:suntocres}
\end{figure*}

\subsection{The new calibration in context}

In Fig.~\ref{f:suntocres} we summarize the ingredients of our calibration by
using the correlation between concentration $c$ and luminosity; well known for some time  (see e.g. Djiorgovski and Meylan 1994, hereinafter DM94). In this
figure, the shape of the symbols represent the different Galactic
sub-populations, while the colours code the different HB morphologies. Symbol
size is proportional to the IQR2 values.

The largest impact is due to the total cluster mass. At any given value of 
concentration there is a clear gradient, the larger is the mass the more
extended is the Na-O anti-correlation, regardless if GCs belong to the
bulge/disc or halo population. Again, this strongly corroborates the idea that
mass is the main driving parameter. The second parameter is the concentration: at
any fixed $M_V$ the IQR decreases as $c$ increases. These trends explain
the observed relations in Fig.~\ref{f:iqrres48} and the dependence on structural
parameters of our resulting empirical calibration. This adds the
`axis' represented by the chemistry of multiple stellar populations to this plot.

\begin{figure}
\centering
\includegraphics[scale=0.40]{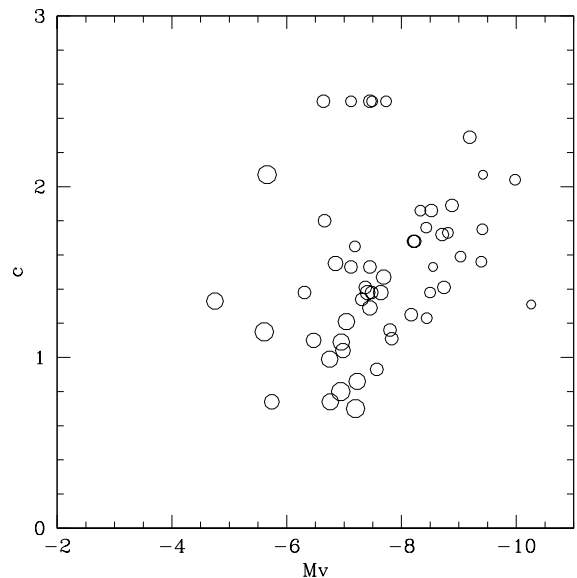}
\caption{Correlation between concentration and luminosity for Galactic GCs,
where are only plotted GCs with fraction of FG stars estimated in Milone et al.
(2017) from $HST$ photometry. Symbol sizes are proportional to the fractions,
in steps of 10\% from about 10 to about 70\%.}
\label{f:cMvN1}
\end{figure}

Photometry may also lend supporting evidence, when the parameters $c$ and $M_V$
are simultaneously considered, as shown in Fig.~\ref{f:cMvN1}. Here we plot
again the $c-M_V$ correlation but this time the symbol sizes are proportional to
the fraction of FG stars N$_1$/N$_{tot}$ estimated from $HST$ photometry by
Milone et al. (2017). Although the sample is now more limited (about the half of
the sample used in the present work) this figure is clearly specular to
Fig.~\ref{f:suntocres}. GCs with larger fractions of FG stars (small fractions
of SG stars) are confined at smaller luminosities (masses) at any given
concentration.

The existence of more extended Na-O anti-correlations in more
loose GCs seem to conflict with the idea that SG stars usually form more centrally
concentrated (but see Gratton et al. 2012 and Bastian and Lardo 2018 for
exceptions). Carretta et al. (2014a) put forward the possibility that low
concentration GCs were originally more massive than present day, but lost more
mass, since at fixed mass the impact of tidal shocks is inversely proportional
to the density of a cluster (Spitzer 1958). This hypothesis is addressed 
in Sect. 4.2. below.

Finally, Fig.~\ref{f:suntocres} is more than a compact summary of our empirical
calibration. As discussed in the pioneering studies by Djorgovski (1991) and
DM94 the correlation between luminosity and concentration also represents
a summary of the initial conditions and the following dynamical evolution of the
GC system.  The concentration parameter can be assumed as a
measure of the dynamical evolutionary stage of clusters since most of these
systems are lead toward higher and higher concentration by the combined impact
of internal evolution and environment-induced effects. Two-body relaxation
pushes stars on the high velocity tail of the Maxwellian velocity distribution,
where they can be more efficiently affected by the Galactic tidal field and
escape the cluster, whose concentration increases (e.g. Murray and Lin 1992,
DM94).  Since at any given $M_V$ concentration and central relaxation time
$t_{rc}$ are anti-correlated, the zone of avoidance at low luminosities and high
$c$ reflects the results of dynamical evolution, which drives low mass clusters
with short $t_{rc}$ toward tidal  disruption or core-collapse (DM94). On the other
hand, the absence of luminous GCs of low concentration is simply explained as a
fossil record of the initial conditions, as no mechanism is able to explain
this second empty region (e.g. Djorgovski  1991).

Bellazzini et al. (1996: B96) attempted to disentangle the relative weight of
evolutionary vs primordial mechanisms in GCs of different galactic
sub-populations by simulating a series of King models in the assumptions that
$c$ and $M_V$ were either correlated or uncorrelated. 
Although their definition of inner and outer GC populations is different from the
one adopted in the present work, by comparing the output of their simulations
with the distribution of real Galactic GCs of various sub-populations they 
concluded that the correlation between concentration and luminosity must be of
primordial origin, being strong for the outer halo GCs, populating regions where
the tidal field is weak, and less clear or absent for inner halo GCs, where it
may have been somewhat mitigated by the effects of dynamical evolution. More
massive GCs are thus born with higher central concentration. 

However, we must not forget that while current masses may be estimated from 
total luminosities or kinematics, the original masses of GCs are usually only
educated guesses, even if the two quantities must be obviously correlated (see
e.g. Kruijssen 2015, Balbinot and Gieles 2018). The greatest unknown is
represented by the dynamical evolution processes required to map the initial
masses in what is currently observed in GCs of the different Galactic
sub-populations.

\subsection{Present and initial masses, and dynamical mass loss}

The IQR2 values show a good correlation with a combination of present-day 
structural parameter $M_V$ and $c$ and
B96 also found the relation between cluster absolute magnitude and 
concentration be mostly primordial in origin. Since $M_V$ is not supposed to 
change very much during the lifetime of a cluster (Murray and Lin 1992), all
together these observations may point out that the impact of the cluster 
dynamical evolution is minimal; just the minimum required to empty the region at
high $c$, faint $M_V$.

This occurrence would seem to be in agreement with the theoretical work by
Reina-Campos  et al. (2018) who addressed the impact of cluster disruption on
multiple populations by deriving statistical estimates of the initial-to-present
mass ratios from E-MOSAICS cosmological simulations.  By using the sample of GCs
with fractions of enriched SG stars $F_{enrich}$ derived from photometry (Milone
et  al. 2017), they concluded that dynamical evolution (tidal shocks and/or
evaporation) cannot explain the rise of this fraction as the cluster mass
increases. Similar arguments are used by Kruijssen (2015) in his
two-phase model, in particular for metal-poor GCs. Reina-Campos et al. (2018)
conclude that dynamical mass loss could not have a large effect on the present
day masses of GCs and the detected fractions as well as the ratios FG/SG must
probably reflect initial  values. Were this the case, a big problem would arise
for the mass-budget limited models, because one of the strongholds more
frequently used to alleviate the so called mass-budget problem is the
preferential loss of FG stars up to 90\% of the cluster mass to simultaneously
account for the scarce  yields of nuclearly polluted matter and the large
observed fraction of SG stars in GCs (see Gratton et al. 2012, Bastian and Lardo
2018 for extensive discussion of this issue).  Usually, in this approach
primordial proto-GCs are assumed to have been much more massive than today,
factors from 10 to 100 are required to eliminate the mass-budget problem, in
stern contrast with the estimates of the initial masses given above. 

The main role  is supposed to be played by the fast mass loss from stellar
evolution: gas  expulsion and ejecta from FG Supernovae cause the cluster to
expand, and stars to escape, reducing the original mass by half in the first Gyr
(Vesperini et al. 2010, Khalaj and Baumgardt 2015, Baumgardt et al. 2019). 
According to most scenarios for multiple populations, at this epoch the Na-O
anti-correlation must be completely developed and  SG stars more centrally
concentrated. As a consequence, the preferential mass loss of  FG stars results
into a dramatic increase of the SG fraction in GCs at early times of their
lifetime. In the next phase, two-body relaxation realizes spatial mixing of the
different population, so that the ensuing dynamical evolution acts more or less
in the same way on different components and the ratio SG/FG stabilizes (e.g.
Vesperini et al. 2010).

We may tackle this issue since our calibration of IQR2 provides sufficiently
large statistics to have a better insight. When observed, best currently
available
proper motions of real Galactic GCs from Gaia DR2 (Gaia Collaboration et al.
2018) are used to compute realistic orbits for the
GCs we may see how the resulting changes in masses compare with the Na-O
anti-correlation.

Baumgardt et al. (2019: BHSB) and Balbinot and Gieles (2018) derived the ratio between
the current masses and the inizial masses of Galactic GCs essentially by
assuming  a mass loss law and playing backward the orbital motions of GCs until
their  primordial masses correctly reproduce the observed present day values. 
While we used the study of Baumgardt et al. (2019) we note that the correlation
between their estimates and those by Balbinot and Gieles (2018) is significant,
at better than 99.9\%.

\begin{figure}
\centering
\includegraphics[scale=0.40]{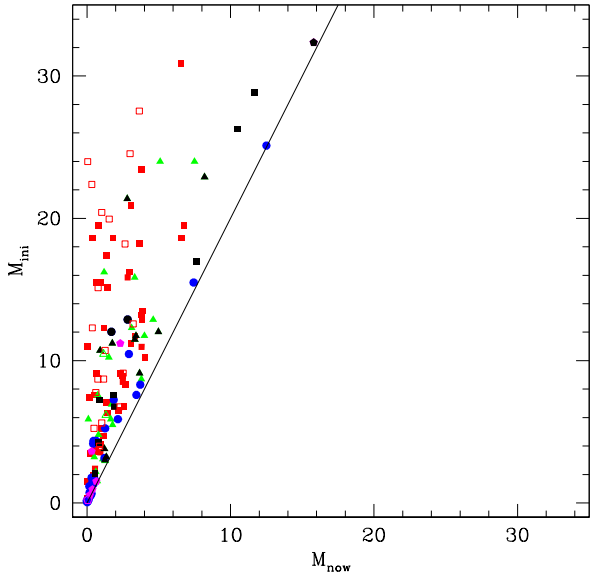}
\caption{Comparison of present and initial GC masses from Baumgardt et al.
(2019). Filled symbols are as in Fig.~\ref{f:cfriqr1}, whereas empty symbols
indicate core-collapse GCs. The line is traced at $M_{ini} = 2 M_{now}$.}
\label{f:mininow}
\end{figure}

The derived initial values are on average $\sim 15$ times the present values
(see Fig.~\ref{f:mininow}), with a clear trend proceding toward the outskirts
of the Galaxy. Bulge/disc GCs were originally 23.1 ($\sigma=80.7$, 76 objects) 
times more massive, on average, while this number decreases to 6.4 times
($\sigma=10.4$, 34 GCs) for inner halo GCs. Outer halo GCs and those 
associated to Sgr were found on average only 4.5 ($\sigma=2.5$, 27 objects)
and 4.3 ($\sigma=3.3$, seven objects) times more massive than the current values,
respectively. No cluster is located rightward the line traced at 
$M_{ini} = 2 M_{now}$, that simply models a 50\% mass loss from stellar
evolution.

\begin{figure}
\centering
\includegraphics[scale=0.40]{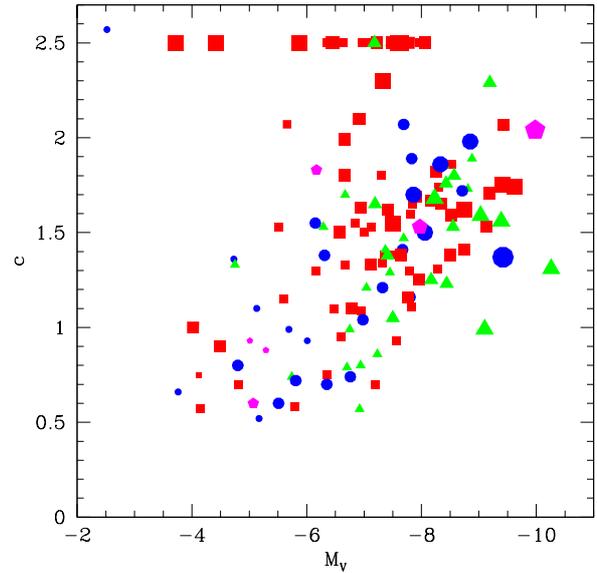}
\caption{Correlation between concentration and luminosity for Galactic GCs.
Different colour and shapes represent the various Galactic populations as in
Fig.~\ref{f:suntocres}. Symbol sizes are proportional to the initial masses
estimated by Baumgardt et al. (2019). In unit of $10^5 M_\odot$ the adopted
ranges for the scaling of sizes are 0-1, 1-10, 10-20, and $>20$.}
\label{f:minigrandi}
\end{figure}

In Fig.~\ref{f:minigrandi} the correlation $c-M_V$ is again represented by
dividing the GCs in various Galactic sub-populations. However, this time the size
of symbols reflects the initial masses as estimated in Baumgardt et al. (2019),
not the present day masses.
Not all the current faint clusters were originally born with small masses.
Classic example are the bulge/disc GCs 1636-283 (ESO452-SC11) and Terzan 4,
present day absolute magnitudes $< -4.5$ and $c \sim 1$. According to Baumgardt
et al. they were born with initial masses comparable to those of M~15 or 47~Tuc
(with $M_V > -9.0$). Their orbital parameters show that these GCs spent almost an
Hubble time near the central regions of the Galaxy, likely affected by (even
severe) tidal shocks. During their lifetimes  the dynamical evolution reduced
their masses to a few per cent of the original  ones. 

Apart from these and other few cases, masses are rather smoothly translated into
the present day absolute magnitudes (masses), with a simple scaling. Thus, the
comparison of Fig.~\ref{f:minigrandi} with Fig.~\ref{f:suntocres} easily 
allows to see that the extent of the Na-O anti-correlation is apparently
already  dictated by the original mass of clusters. Baumgardt et al. (2019) also
found for the least dynamically evolved  GCs a slope of the mass function about
1 dex higher than a Kroupa mass function in the range 0.2 to 0.8 $M_\odot$. The
authors considered this result as an evidence that GC started with a bottom
light mass function. In this case, due to the fewer low-mass stars present, a
larger fraction of stars may be polluted, partly alleviating the tension between
the expected and observed SG fractions.

\begin{figure}
\centering
\includegraphics[scale=0.40]{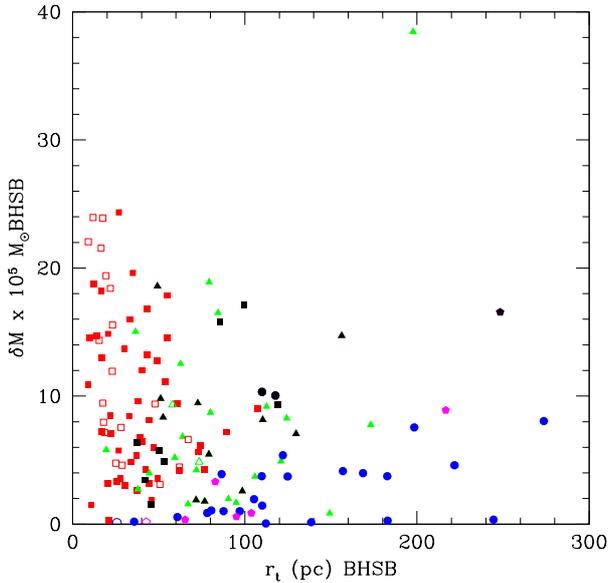}
\caption{Mass loss from Baumgardt et al. (2019), in unit of $10^5 M_\odot$, 
as a function of the tidal radius for Galactic GCs. Symbols are as
Fig.~\ref{f:mininow}.}
\label{f:dmrt}
\end{figure}

Finally, in Fig.~\ref{f:dmrt} we plot the difference between initial and present
day masses (Baumgardt et al. 2019, and private communication) as a function of
the tidal radius in pc from the fit of N-body
models\footnote{https://people.smp.uq.edu.au/HolgerBaumgardt/globular/}. No
evidence of extended GCs losing more mass than the compact clusters is found, as 
advanced instead in Carretta et al. (2014a) to explain more extended
anti-correlations in more loose GCs.

Remarkably, the conclusion from this section is that the present-day masses of GCs seem to
preserve essentially the ranking provided by the initial masses, apart from a
few exception. When the observed IQR[O/Na] values for our 22 calibrators are
correlated either to $M_V$ or $\log M_{ini}$ we derive the same value of the
Pearson correlation coefficient ($r_P=-0.48$ and $r_P=0.48$ respectively). The
bottom line is that both correlations are equally significant, at a confidence
level greater than 99.9\%.

However, taking into account the number of assumptions still entering the
computation of the initial GC masses (and we cannot know their original
concentration), we stand for a calibration including the
present day masses, as expressed by the proxy of total luminosities, a rather
simple quantity to be measured. In addition, our empirical calibration offers a
precious tool with which to examine globular clusters in nearby galaxies putting them on
the same scale and allowing a meaningful comparison with GCs of the Galactic
system. This will be the subject of a forthcoming paper.

\section{Summary and future work}

The extent of the Na-O anti-correlation observed in GCs is an useful quantitative
measure of the impact of multiple population phenomenon in a globular cluster.
Obtaining this measure requires a large investment of observing time even with 
high multiplex spectrographs at 8-10 m class telescopes.

We the used 22 GCs from our unprecedented homogeneous FLAMES survey to calibrate
observed IQR[O/Na] values in term of structural parameters and HB morphology. We
find that the combination of total absolute cluster magnitude (a proxy for the
cluster present-day mass) and concentration is very well correlated to the
extent of the Na-O anti-correlation in GCs. When a term describing the HB
morphology is added, our empirical calibration returns values on the identity
line with respect to the observed values on our system, based on more than 2000
red giant in 22 GCs.

We were able to provide empirical estimates of the Na-O anti-correlation in 95
Galactic GCs. When added to the 25 GCs observed in our 
FLAMES survey, we can homogeneously provide quantitative estimates of the
multiple population phenomenon in a large sample of 120 GCs (80\% of all GCs in
the Milky Way). The calibration cannot be applied to core-collapse GCs, but
direct observation of Na-O abundances in at least three GCs show that the
anti-correlation is preserved even after the core-collapse.

The large number of data points  allow us to uncover that the main differences in the extent of
the multiple population phenomenon are a direct reflection of the mean
properties of the parent cluster sub-population (outer and inner halo, disc) in
the Galaxy (luminosity function, distribution of sizes and concentration). This
accounts for the findings by Carretta et al. (2010a) that the chemical
signatures of multiple populations included a contribution due to the location
in the Galaxy.

Galactic clusters that feature a large Na-O anti-correlation are segregated along the
correlation between $c$ and M$_V$, at high luminosities (mass) 
values at each given concentration. This well known correlation explains the 
dependence of our calibrated IQR2 on structural parameters.
A comparison with recent estimates of initial masses suggests that the extension
of Na-O anti-correlation is already dependent on the original GC masses. The
following dynamical evolution seems essentially to preserve this dependence
through a scaling of the masses to the present-day values, apart a few
exceptions (mostly GCs spending most of their lifetimes in the central regions
of the Galaxy, where the effects of tidal interaction are strong).
We thus recommend to adopt the present calibration using the observed total
absolute magnitudes $M_V$ that are easily obtained from the observations and
are not affected by assumptions regarding the estimates of initial GC masses.

Moreover, this calibration can be easily applied to GC systems in nearby
external galaxies: a forthcoming paper will be devoted to this.
A first check is already provided by GCs in our Galaxy that are associated to
the closest external galaxy, the dwarf spheroidal Sgr (Ibata et al. 1994).
The disruption of this dSph released in the Milky Way a number of GCs, all
listed in the Harris (1996) catalogue, and in Table~\ref{t:tabappGC}.
According to our calibration the properties of these migrant GCs cannot be 
distinguished from autochthonous Milky Way GCs in this context.
We conclude that our empirical calibration is a useful compact way to represent
and quantify the properties of multiple populations in GCs.

\begin{acknowledgements}
I would like to thank the referee for a constructive review.
I wish to thank Angela Bragaglia for valuable help and suggestions, Emanuele
Dalessandro for useful discussions, Antonio Sollima and Holger Baumgardt for 
providing their estimates of initial GC masses, and Raffaele Gratton who first 
pointed out the role of the concentration.
This research has made use of the SIMBAD database (in particular  Vizier),
operated at CDS, Strasbourg, France, of the NASA's Astrophysical Data System.
\end{acknowledgements}

\begin{appendix}

\section{Derived values for IQR2}
In the following table we list the relevant parameters and the results from the
calibration of IQR2 (Eq. 2).
The values for total absolute magnitude $M_V$, concentration $c$, metallicity
[Fe/H] are from Harris (1996, 2010 on line edition). The Galactic population is
the same as in Table~1 and is adopted from Carretta et al. (2010a), whereas the
HBR values are from mackey and van den Bergh (2005) except for Terzan 8, whose
HB index is taken from Carretta et al. (2010a).

Note that the values of IQR2 are actually the observed IQR[O/Na] values for the 22
calibrating GCs in Table~1, as well as for the three core-collapse GCs NGC~6397,
NGC~6752, and NGC~7099 (M~30). These three GCs are studied in our FLAMES survey,
but not used to derive the calibration for IQR2.

\begin{table*}
\centering
\caption{Relevant parameters and calibrated IQR2 values for Galactic GCs}
\begin{tabular}{lcccccr}
\hline
GC Name          & $M_V$  &  c   & [Fe/H]  &   IQR2   &    t    &     HBR \\
\hline
NGC104 47 Tuc    &  -9.42 & 2.07 &  -0.72  &   0.472  &    0    &   -0.99  \\
NGC288           &  -6.75 & 0.99 &  -1.32  &   0.776  &    1    &    0.98  \\
NGC362           &  -8.43 & 1.76 &  -1.26  &   0.644  &    1    &   -0.87  \\
NGC1261          &  -7.80 & 1.16 &  -1.27  &   0.663  &    2    &   -0.71  \\
ERIDANUS         &  -5.13 & 1.10 &  -1.43  &   0.133  &    2    &   -1.00  \\
PAL2             &  -7.97 & 1.53 &  -1.42  &   0.630  &    3    &   -0.10  \\
NGC1851          &  -8.33 & 1.86 &  -1.18  &   0.693  &    2    &   -0.32  \\
NGC1904  M79     &  -7.86 & 1.70 &  -1.60  &   0.759  &    2    &    0.89  \\
NGC2298          &  -6.31 & 1.38 &  -1.92  &   0.484  &    2    &    0.93  \\
NGC2419          &  -9.42 & 1.37 &  -2.15  &   1.083  &    2    &    0.86  \\
NGC2808          &  -9.39 & 1.56 &  -1.14  &   0.999  &    1    &   -0.49  \\
PAL3             &  -5.69 & 0.99 &  -1.63  &   0.341  &    2    &   -0.50  \\
NGC3201          &  -7.45 & 1.29 &  -1.59  &   0.634  &    1    &    0.08  \\
PAL4             &  -6.01 & 0.93 &  -1.41  &   0.367  &    2    &   -1.00  \\
NGC4147          &  -6.17 & 1.83 &  -1.80  &   0.259  &    3    &    0.66  \\
NGC4372          &  -7.79 & 1.30 &  -2.17  &   0.809  &    0    &    1.00  \\
RUP106           &  -6.35 & 0.70 &  -1.68  &   0.539  &    2    &   -0.82  \\
NGC4590  M68     &  -7.37 & 1.41 &  -2.23  &   0.372  &    1    &    0.17  \\
NGC4833          &  -8.17 & 1.25 &  -1.85  &   0.945  &    1    &    0.93  \\
NGC5024  M53     &  -8.71 & 1.72 &  -2.10  &   0.810  &    2    &    0.81  \\
NGC5053          &  -6.76 & 0.74 &  -2.27  &   0.761  &    2    &    0.52  \\
NGC5272  M3      &  -8.88 & 1.89 &  -1.50  &   0.695  &    1    &    0.08  \\
NGC5286          &  -8.74 & 1.41 &  -1.69  &   0.930  &    0    &    0.80  \\
NGC5466          &  -6.98 & 1.04 &  -1.98  &   0.699  &    2    &    0.58  \\
NGC5634          &  -7.69 & 2.07 &  -1.88  &   0.494  &    2    &    0.91  \\
NGC5694          &  -7.83 & 1.89 &  -1.98  &   0.599  &    2    &    1.00  \\
IC4499           &  -7.32 & 1.21 &  -1.53  &   0.647  &    2    &    0.11  \\
NGC5824          &  -8.85 & 1.98 &  -1.91  &   0.739  &    2    &    0.79  \\
PAL5             &  -5.17 & 0.52 &  -1.41  &   0.426  &    2    &   -0.40  \\
NGC5897          &  -7.23 & 0.86 &  -1.90  &   0.847  &    1    &    0.86  \\
NGC5904  M5      &  -8.81 & 1.73 &  -1.29  &   0.741  &    1    &    0.31  \\
NGC5927          &  -7.81 & 1.60 &  -0.49  &   0.468  &    0    &   -1.00  \\
NGC5986          &  -8.44 & 1.23 &  -1.59  &   0.958  &    1    &    0.97  \\
Lynga7           &  -6.60 & 0.95 &  -1.01  &   0.474  &    0    &   -1.00  \\
PAL14            &  -4.80 & 0.80 &  -1.62  &   0.180  &    2    &   -1.00  \\
NGC6093  M80     &  -8.23 & 1.68 &  -1.75  &   0.784  &    1    &    0.93  \\
NGC6101          &  -6.94 & 0.80 &  -1.98  &   0.811  &    1    &    0.84  \\
NGC6121  M4      &  -7.19 & 1.65 &  -1.16  &   0.373  &    1    &   -0.06  \\
NGC6144          &  -6.85 & 1.55 &  -1.76  &   0.534  &    0    &    1.00  \\
NGC6139          &  -8.36 & 1.86 &  -1.65  &   0.702  &    1    &    0.91  \\
TER3             &  -4.82 & 0.70 &  -0.74  &   0.221  &    0    &   -1.00  \\
NGC6171  M107    &  -7.12 & 1.53 &  -1.02  &   0.522  &    0    &   -0.73  \\
1636-283         &  -4.02 & 1.00 &  -1.50  &   0.025  &    0    &   -0.40  \\
NGC6205  M13     &  -8.55 & 1.53 &  -1.53  &   0.868  &    1    &    0.97  \\
NGC6218  M12     &  -7.31 & 1.34 &  -1.37  &   0.863  &    0    &    0.97  \\
NGC6229          &  -8.06 & 1.50 &  -1.47  &   0.699  &    2    &    0.24  \\
NGC6235          &  -6.29 & 1.53 &  -1.28  &   0.420  &    1    &    0.89  \\
NGC6254  M10     &  -7.48 & 1.38 &  -1.56  &   0.565  &    0    &    0.98  \\
PAL15            &  -5.51 & 0.60 &  -2.07  &   0.626  &    2    &    1.00  \\
NGC6266  M62     &  -9.18 & 1.71 &  -1.18  &   0.848  &    0    &    0.32  \\
NGC6273  M19     &  -9.13 & 1.53 &  -1.74  &   0.979  &    0    &    0.96  \\
NGC6287          &  -7.36 & 1.38 &  -2.10  &   0.694  &    0    &    0.98  \\
NGC6304          &  -7.30 & 1.80 &  -0.45  &   0.295  &    0    &   -1.00  \\
NGC6316          &  -8.34 & 1.65 &  -0.45  &   0.553  &    0    &   -1.00  \\
NGC6341  M92     &  -8.21 & 1.68 &  -2.31  &   0.740  &    1    &    0.91  \\
NGC6333  M9      &  -7.95 & 1.25 &  -1.77  &   0.844  &    0    &    0.87  \\
NGC6356          &  -8.51 & 1.59 &  -0.40  &   0.608  &    0    &   -1.00  \\
NGC6352          &  -6.47 & 1.10 &  -0.64  &   0.393  &    0    &   -1.00  \\
IC1257           &  -6.15 & 1.55 &  -1.70  &   0.399  &    2    &    1.00  \\
NGC6366          &  -5.74 & 0.74 &  -0.59  &   0.388  &    1    &   -0.97  \\
NGC6362          &  -6.95 & 1.09 &  -0.99  &   0.539  &    0    &   -0.58  \\
TER4     HP4     &  -4.48 & 0.90 &  -1.41  &   0.315  &    0    &    1.00  \\
LILLER1          &  -7.32 & 2.30 &  -0.33  &   0.114  &    0    &   -1.00  \\
\hline
\end{tabular}
\end{table*}

\addtocounter{table}{-1}

\begin{table*}
\centering
\caption{continue}
\begin{tabular}{lcccccr}
\hline
GC Name          & $M_V$  &  c   & [Fe/H]  &   IQR2   &    t    &     HBR \\
\hline
NGC6380  TON1    &  -7.50 & 1.55 &  -0.75  &   0.427  &    0    &   -1.00  \\
Ton2             &  -6.17 & 1.30 &  -0.70  &   0.261  &    0    &   -1.00  \\
NGC6388          &  -9.41 & 1.75 &  -0.55  &   0.644  &    0    &   -1.00  \\
NGC6402  M14     &  -9.10 & 0.99 &  -1.28  &   1.137  &    1    &    0.65  \\
NGC6401          &  -7.90 & 1.69 &  -1.02  &   0.610  &    0    &    0.35  \\
NGC6397          &  -6.64 & 2.50 &  -2.02  &   0.274$^a$ &    0 &    0.98  \\
PAL6             &  -6.79 & 1.10 &  -0.91  &   0.455  &    0    &   -1.00  \\
NGC6426          &  -6.67 & 1.70 &  -2.15  &   0.395  &    1    &    0.58  \\
TER5     Ter11   &  -7.42 & 1.62 &  -0.23  &   0.385  &    0    &   -1.00  \\
NGC6440          &  -8.75 & 1.62 &  -0.36  &   0.643  &    0    &   -1.00  \\
NGC6441          &  -9.63 & 1.74 &  -0.46  &   0.660  &    0    &   -1.00  \\
UKS1             &  -6.91 & 2.10 &  -0.64  &   0.109  &    0    &   -1.00  \\
NGC6496          &  -7.20 & 0.70 &  -0.46  &   0.683  &    0    &   -1.00  \\
Djorg2           &  -7.00 & 1.50 &  -0.65  &   0.348  &    0    &   -1.00  \\
NGC6517          &  -8.25 & 1.82 &  -1.23  &   0.662  &    0    &    0.62  \\
TER10            &  -6.35 & 0.75 &  -1.79  &   0.499  &    0    &   -1.00  \\
NGC6535          &  -4.75 & 1.33 &  -1.79  &   0.208  &    1    &    1.00  \\
NGC6528          &  -6.57 & 1.50 &  -0.11  &   0.265  &    0    &   -1.00  \\
NGC6539          &  -8.29 & 1.74 &  -0.63  &   0.510  &    0    &   -1.00  \\
NGC6544          &  -6.94 & 1.63 &  -1.40  &   0.522  &    0    &    1.00  \\
NGC6541          &  -8.52 & 1.86 &  -1.81  &   0.744  &    0    &    1.00  \\
NGC6553          &  -7.77 & 1.16 &  -0.18  &   0.623  &    0    &   -1.00  \\
IC1276   PAL7    &  -6.67 & 1.33 &  -0.75  &   0.347  &    0    &   -1.00  \\
TER12            &  -4.14 & 0.57 &  -0.50  &   0.137  &    0    &   -1.00  \\
NGC6569          &  -8.28 & 1.31 &  -0.76  &   0.688  &    0    &   -0.82  \\
NGC6584          &  -7.69 & 1.47 &  -1.50  &   0.592  &    1    &   -0.15  \\
NGC6626  M28     &  -8.16 & 1.67 &  -1.32  &   0.732  &    0    &    0.90  \\
NGC6638          &  -7.12 & 1.33 &  -0.95  &   0.516  &    0    &   -0.30  \\
NGC6637  M69     &  -7.64 & 1.38 &  -0.64  &   0.517  &    0    &   -1.00  \\
NGC6642          &  -6.66 & 1.99 &  -1.26  &   0.213  &    0    &   -0.04  \\
NGC6652          &  -6.66 & 1.80 &  -0.81  &   0.171  &    0    &   -1.00  \\
NGC6656  M22     &  -8.50 & 1.38 &  -1.70  &   0.907  &    0    &    0.91  \\
PAL8             &  -5.51 & 1.53 &  -0.37  &   0.048  &    0    &   -1.00  \\
NGC6712          &  -7.50 & 1.05 &  -1.02  &   0.656  &    1    &   -0.62  \\
NGC6715  M54     &  -9.98 & 2.04 &  -1.49  &   1.169  &    3    &    0.54  \\
NGC6717  PAL9    &  -5.66 & 2.07 &  -1.26  &   0.109  &    0    &    0.98  \\
NGC6723          &  -7.83 & 1.11 &  -1.10  &   0.761  &    0    &   -0.08  \\
NGC6749          &  -6.70 & 0.79 &  -1.60  &   0.787  &    1    &    1.00  \\
NGC6752          &  -7.73 & 2.50 &  -1.54  &   0.772$^a$ &    0 &    1.00  \\
NGC6760          &  -7.84 & 1.65 &  -0.40  &   0.456  &    0    &   -1.00  \\
NGC6779  M56     &  -7.41 & 1.38 &  -1.98  &   0.704  &    1    &    0.98  \\
TER7             &  -5.01 & 0.93 &  -0.32  &   0.173  &    3    &   -1.00  \\
PAL10            &  -5.79 & 0.58 &  -0.10  &   0.454  &    0    &   -1.00  \\
ARP2             &  -5.29 & 0.88 &  -1.75  &   0.425  &    3    &    0.53  \\
NGC6809  M55     &  -7.57 & 0.93 &  -1.94  &   0.725  &    0    &    0.87  \\
TER8             &  -5.07 & 0.60 &  -2.16  &   0.541  &    3    &    1.00$^b$  \\
PAL11            &  -6.92 & 0.57 &  -0.40  &   0.677  &    1    &   -1.00  \\
NGC6838  M71     &  -5.61 & 1.15 &  -0.78  &   0.257  &    0    &   -1.00  \\
NGC6864  M75     &  -8.57 & 1.80 &  -1.29  &   0.650  &    1    &   -0.07  \\
NGC6934          &  -7.45 & 1.53 &  -1.47  &   0.570  &    1    &    0.25  \\
NGC6981  M72     &  -7.04 & 1.21 &  -1.42  &   0.596  &    1    &    0.14  \\
NGC7006          &  -7.67 & 1.41 &  -1.52  &   0.596  &    2    &   -0.28  \\
NGC7078  M15     &  -9.19 & 2.29 &  -2.37  &   0.501  &    1    &    0.67  \\
NGC7089  M2      &  -9.03 & 1.59 &  -1.65  &   0.933  &    1    &    0.92  \\
NGC7099  M30     &  -7.45 & 2.50 &  -2.27  &   0.607$^a$ &    1 &    0.89  \\
PAL13            &  -3.76 & 0.66 &  -1.88  &   0.124  &    2    &   -0.20  \\
NGC7492          &  -5.81 & 0.72 &  -1.78  &   0.618  &    2    &    0.81  \\
\hline
\end{tabular}
\begin{list}{}{}
\item[(a)] not used in the calibrating Eq. (2).
\item[(b)] HBR from Carretta et al. (2010a)
\end{list}
\label{t:tabappGC}
\end{table*}

\end{appendix}


\begin{thebibliography}{}

\bibitem[]{} Balbinot, E., Gieles, M. 2018, MNRAS, 484, 2479 
\bibitem[]{} Bastian, N., Lardo, C. 2018, ARA\&A, 56, 83
\bibitem[]{} Baumgardt, H., Hilker, M., Sollima, A., Bellini, A. 2019, 
 MNRAS, 482, 5138 
\bibitem[]{} Bellazzini, M., Vesperini, E., Ferraro, F.R., Fusi Pecci, F. 1996,
 MNRAS, 279, 337 
\bibitem[]{} Bellazzini, M., Ibata, R.A., Chapman, S.C. et al. 2008, , AJ, 136, 
 1147 
\bibitem[]{} Bragaglia, A., Carretta, E., Sollima, A. et al. 2015, 
 A\&A, 583, A69 
\bibitem[]{} Briley, M.M. 1997, AJ, 114, 1051 
\bibitem[]{} Briley, M.M., Smith, G.H., Claver, C.F. 2001, AJ, 122, 2561 
\bibitem[]{} Carretta, E. 2006, AJ, 131, 1766 
\bibitem[]{} Carretta, E. 2015, ApJ, 810, 148 
\bibitem[]{} Carretta, E. 2016, arXiv:1611.04728 
\bibitem[]{} Carretta, E., Bragaglia, A. 2018, A614, A109 
\bibitem[]{} Carretta, E., Bragaglia, A., Gratton R.G. et al. 2006, A\&A, 
  450, 523 
\bibitem[]{} Carretta, E., Recio-Blanco, A., Gratton, R.G., Piotto, G.,
  Bragaglia, A. 2007a, ApJ, 671, L125 
\bibitem[]{} Carretta, E., Bragaglia, A., Gratton, R.G. et al. 2007b, 
 A\&A, 464, 939 
\bibitem[]{} Carretta, E., Bragaglia, A., Gratton, R.G., Lucatello,
S. Momany, Y. 2007c, A\&A, 464, 927 
\bibitem[]{}  Carretta, E., Bragaglia, A., Gratton, R.G. et al. 2009a, 
  A\&A, 505, 117  
\bibitem[]{} Carretta, E., Bragaglia, A., Gratton, R.G., Lucatello, S. 2009b, 
 A\&A, 505, 139 
\bibitem[]{} Carretta, E., Bragaglia, A., Gratton, R.G. et al. 2010a, 
 A\&A, 516, 55 
\bibitem[]{} Carretta, E., Bragaglia, A., Gratton, R.G. et al. 2010b, A\&A, 520,
 95 
\bibitem[]{} Carretta, E., Bragaglia, A., Gratton, R.G. et al. 2010c, ApJ, 714, 
 L7 
\bibitem[]{} Carretta, E., Lucatello, S., Gratton, R.G., Bragaglia, A., D'Orazi,
  V. 2011, A\&A, 533, 69 
\bibitem[]{} Carretta, E., Bragaglia, A., Gratton, R.G., Lucatello, S., 
  D'Orazi, V. 2012, ApJ, 750, L14 
\bibitem[]{} Carretta, E., Bragaglia, A., Gratton, R.G. et al. 2013, A\&A, 557, 
  A138 
\bibitem[]{} Carretta, E., Bragaglia, A., Gratton, R.G. et al. 2014a, A\&A, 
  564, A60 
\bibitem[]{} Carretta, E., Bragaglia, A., Gratton, R.G. et al. 2014b, A\&A, 
 561, A87 
\bibitem[]{} Carretta, E., Bragaglia, A., Gratton, R.G. et al. 2015, 
 A\&A, 578, A116 
\bibitem[]{} Carretta, E., Bragaglia, A., Lucatello, S. et al. 2018, A\&A, 615, 
 A17 
\bibitem[]{} Cassisi, S., Castellani, V., degl'Innocenti, S., Salaris, M.,
 Weiss, A. 1999, A\&AS, 134, 103 
\bibitem[]{} Dalessandro, E., Schiavon, R.P., Rood, R.T. et al. 2012, AJ, 144, 
 126 
\bibitem[]{} D'Antona, F., Caloi, V., Montalbán, J., Ventura, P., Gratton, R.
  2002, A\&A, 395, 69 
\bibitem[]{} D'Antona, F., Vesperini, E., D'Ercole, A. et al. 2016, MNRAS, 458, 
 2122 
\bibitem[]{} Decressin, T., Meynet, G., Charbonnel C. Prantzos, N.,
 Ekstrom, S. 2007, A\&A, 464, 1029 
\bibitem[]{} Denisenkov, P.A.,\&  Denisenkova, S.N. 1989, A.Tsir., 1538, 11
\bibitem[]{} D'Ercole, A., Vesperini, E., D'Antona, F., McMillan, S.L.W.,
  Recchi, S. 2008, MNRAS, 391, 825 
\bibitem[]{} Djorgovski, S. 1991, ASPC, 13, 112 
\bibitem[]{} Djorgovski, S., Meylan, G. 1994, AJ, 108, 1292 
\bibitem[]{} Dotter, A., Sarajedini, A., Anderson, J. et al. 2010, ApJ, 708, 698 
\bibitem[]{} Dupree, A.K., Strader, J., Smith, G.H. 2011, ApJ, 728, 155 
\bibitem[]{} Fusi Pecci, F. et al. 1993, AJ, 105, 1145 
\bibitem[]{} Gaia Collaboration et al. 2018, A\&A, 616, A12 
\bibitem[]{} Gratton, R.G., Sneden, C., Carretta, E. 2004, ARA\&A, 42, 385
\bibitem[]{} Gratton, R.G., Carretta, E., Bragaglia, A. 2012, A\&ARv, 20, 50 
\bibitem[]{} Gratton, R.G., Lucatello, S., Bragaglia, A. et al. 2006,  
 A\&A, 455, 271 
\bibitem[]{} Gratton, R.G., Lucatello, S., Bragaglia, A. et al. 2007, A\&A, 
  464, 953 
\bibitem[]{} Gratton, R.G., Carretta, E., Bragaglia, A., Lucatello, S., 
  D'Orazi, V. 2010, A\&A, 517, 81 
\bibitem[]{} Halford, M., Zaritsky, D. 2015, ApJ, 815, 86 
\bibitem[]{} Harris, W.~E. 1996, AJ, 112, 1487
\bibitem[]{} Ibata, R.A., Irwin, M.J., Gilmore, G. 1994, Nature, 370, 194 
\bibitem[]{} Johnson, C.I., Rich, M.R., Pilachowski, C.A. et al. 2015, AJ, 150, 
 63 
\bibitem[]{} Kayser, A., Hilker, M., Grebel, E.K., Willemsen, P.G. 2008, A\&A,
  486, 437 
\bibitem[]{} Khalaj, P., Baumgardt, H. 2015, MNRAS, 452, 924 
\bibitem[]{} King, I.R. 1966, AJ, 71, 64 
\bibitem[]{} Kraft, R.~P. 1994, PASP, 106, 553 
\bibitem[]{} Kruijssen, J.M.D. 2015, MNRAS, 454, 1658 
\bibitem[]{} Langer, G.E., Hoffman, R., Sneden, C. 1993, PASP, 105, 301 
\bibitem[]{} Leaman, R., VandenBerg, D.A., Mendel, J.T. 2013, MNRAS, 436, 122 
\bibitem[]{} Lee, Y.W. 1989, PhD thesis, Yale University
\bibitem[]{} Lee, Y.W. 1990, ApJ, 363, 159
\bibitem[]{} Lee, Y.-W., Gim, H.B., Casetti-Dinescu, D.I. 2007, ApJ, 661, L49
\bibitem[]{} Mackey, A.D., Gilmore, G.F. 2004, MNRAS, 355, 504 
\bibitem[]{} Mackey, A.D., van den Bergh, S. 2005, MNRAS, 360, 631
\bibitem[]{} Marino, A.F., Sneden, C., Kraft, R.P. et al. 2011, A\&A, 532, A8
\bibitem[]{} Marino, A.F., Milone, A.P., Karakas, A.I. et al. 2015, MNRAS, 450,
 815 
\bibitem[]{} Milone, A., Piotto, G., Bedin, L. et al. 2012a, ApJ, 744, 58 
\bibitem[]{} Milone, A.P., Piotto, G., Bedin, L. et al. 2012b, A\&A, 537, A77 
\bibitem[]{} Milone, A.P., Marino, A.F., Dotter et al. 2014, ApJ, 785, 21 
\bibitem[]{} Milone, A.P., Piotto, G., Renzini, A. et al. 2017, MNRAS, 464,
  3636 
\bibitem[]{} Milone, A.P., Marino, A.F., Renzini, A. et al. 2018, MNRAS,
 481, 5098 
\bibitem[]{} Monelli, M., Milone, A.P., Stetson, P.B. et al. 2013, MNRAS, 431,
  2126 
\bibitem[]{} Murray, S.D., Lin, D.N.C. 1992, ApJ, 400, 265 
\bibitem[]{} Norris, J. 1987, ApJ, 313, L65 
\bibitem[]{} Odenkirchen, M., Grebel, E.K., Rockosi, C.M. et al. 2001, ApJ, 548,
 L165 
\bibitem[]{} Odenkirchen, M., Grebel, E.K., Dehnen, W. et al. 2003, AJ, 126,
 2385 
\bibitem[]{} Pancino, E., Rejkuba, M., Zoccali, M., Carrera, R. 2010, A\&A, 524,
  A44 
\bibitem[]{} Pasquini, L., Mauas, P., K\'aufl, H. U., Cacciari, C. 2011, A\&A, 
  531, 35 
\bibitem[]{} Perina, S., Bellazzini, M., Buzzoni, A. et al. 2012, A\&A, 546, 31
\bibitem[]{} Recio-Blanco, A., Aparicio, A., Piotto, G., De Angeli, F., 
 Djorgovski, S.G. 2006, A\&A, 452, 875
\bibitem[]{} Reina-Campos, M., Kruijssen, J.M.D., Pfeffer, J., Bastian, N.,
  Crain, R.A. 2018, MNRAS, 481, 2851 
\bibitem[]{} Saracino, S., Dalessandro, E., Ferraro, F.R. et al. 2015, ApJ, 806,
 152 
\bibitem[]{} Sarajedini, A., Bedin, L.R., Chaboyer, B. et al. 2007, AJ, 133,
 1658 
\bibitem[]{} Simpson, J.D., De Silva, G., Martell, S.L., Navin, C.A., 
 Zucker, D.B. 2017, MNRAS, 472, 2856 
\bibitem[]{} Smith, G.H. 1987, PASP, 99, 67 
\bibitem[]{} Smith, G.H. 2015, PASP, 127, 825
\bibitem[]{} Smith, G.H., Sneden, C., Kraft, R.P. 2002, AJ, 123, 1502 
\bibitem[]{} Smith, G.H., Modi, P.N., Hamren, K. 2013, PASP, 125, 1287
\bibitem[]{} Spitzer, Jr. L. 1958, ApJ, 127, 17 
\bibitem[]{} Ventura, P. D'Antona, F., Mazzitelli, I., Gratton, R. 2001,
  ApJ, 550, L65 
\bibitem[]{} Vesperini, E., McMillan, S.L.W., D'Antona, F., D'Ercole, A. 2010,
 ApJ, 718, L112 
\bibitem[]{} Zinn, R.J. 1986, in Stellar populations, eds. C.A. Norman, A. Renzini, M. Tosi
(Cambridge Uiversity Press, Cambridge), p.73 

\end{thebibliography}
\end{document}